\documentclass[aps, superscriptaddress, prd, floatfix, 10pt]{revtex4}

\usepackage{amsmath, amssymb, amsfonts}
\usepackage{graphicx}
\usepackage{color}
\usepackage{hyperref}

\newcommand{\va}[0]{\vec{A}}
\newcommand{\vp}[0]{\vec{p}}
\newcommand{\vn}[0]{\vec{n}}
\newcommand{\figref}[1]{Fig.\,\ref{#1}}
\newcommand{\tabref}[1]{Table.\,\ref{#1}}

\newcommand{\sgn}{\operatorname{sgn}}

\begin{document}
\title{A ``Helium Atom'' of Space: Dynamical Instability of the Isochoric Pentahedron}

\author{C.~E.~Coleman-Smith} \email{cec24@phy.duke.edu}
\author{B.~M\"uller} \email{muller@phy.duke.edu}
\affiliation{Department of Physics, Duke University, Durham, NC 27708-0305, USA}
\date{\today}

\begin{abstract}
  We present an analysis of the dynamics of the equifacial pentahedron on the
  Kapovich-Millson phase space under a volume preserving Hamiltonian. The classical
  dynamics of polyhedra under such a Hamiltonian may arise from the classical limit of the
  node volume operators in loop quantum gravity. The pentahedron is the simplest
  nontrivial polyhedron for which the dynamics may be chaotic. We consider the
  distribution of polyhedral configurations throughout the space and find indications that
  the borders between certain configurations act as separatrices. We examine the local
  stability of trajectories within this phase space and find that locally unstable regions
  dominate although extended stable regions are present. Canonical and microcanonical
  estimates of the Kolmogorov-Sinai entropy suggest that the pentahedron is a strongly
  chaotic system. The presence of chaos is further suggested by calculations of
  intermediate time Lyapunov exponents which saturate to non zero values.
\end{abstract}

\maketitle

\section{Introduction}

It has long been known that black holes act as thermodynamic systems whose entropy
is proportional to the area of their horizon \cite{Bekenstein:1973ur} and a temperature
that is inversely proportional to their mass \cite{Hawking:1974sw}. This has raised the
question how quickly and how perfectly a black hole thermalizes and thereby effectively
destroys any injected information. Sekino and Susskind have argued that black holes 
are ``fast scramblers'' \cite{Sekino:2008he,Susskind:2011ap}, i.~e.\ systems that render
information practically unretrievable at the maximal possible rate. 

Scrambling information among different degrees of freedom is one particular form
of deterministic chaos. A well known realization of scrambling is provided by the
Baker's map. Other examples of dynamical chaos generated by the nonlinearity
of Einstein's field equations have been studied in the context of various analytical
solutions of general relativity (see \cite{Matinyan:2000zq, Matinyan:2006yc} for a review). The results
of these investigations suggest that dynamical chaos, and thus the tendency to 
lose initially known information effectively irretrievably, is a generic property of
classical gravitation.

These considerations apply to macroscopic gravitational fields, which can be considered 
as thermodynamic systems with a macroscopic number of degrees of freedom. But what
about microscopic black holes with masses near the Planck mass, which possess only
a small number of degrees of freedom? Is there a smallest black hole that can be 
considered as a thermal system? Which mechanism drives the apparent thermal 
equilibration of black holes at the microscopic level? The pursuit of these questions 
requires a quantum theory of gravity. While the true quantum mechanical foundation 
of gravity is still unknown, there are at least two widely explored candidates for such 
a theory: superstring theory \cite{Green:1987sp, Becker:2007} 
and Loop Quantum Gravity \cite{Ashtekar:1987gu, Rovelli:2004tv, Thiemann:2007zz}. Here we consider the problem of the 
microscopic origin of the thermal properties of space-time in the framework of Loop
Quantum Gravity (LQG). 

LQG is an attempt to reconcile general relativity and quantum field theory 
\cite{Rovelli:1989za, Rovelli:1995ac, Rovelli:2010bf,Rovelli:2004tv}, the structure of space-time
emerges naturally from the dynamics of a graph of $SU(2)$ spins. Naively the nodes of
this graph can be thought of as representing granules of space-time, the spins connecting
these nodes can be thought of as the faces of these granules. The volume of these
granules, along with the areas of the connected faces are naturally quantized
\cite{Rovelli:1994ge}. 

There has recently been a focus on finding a semi-classical description of the spectrum of
the volume operator at one of these nodes. There have been several reasonable candidates
for the quantum volume operator (cites), a semi-classical limit may serve to pick out a
particular one of these forms. The volume preserving deformation of polyhedra has recently
emerged as a candidate for this semi-classical limit \cite{Rovelli:2006fw,Baez:1999tk}. In
this scheme the black hole thermodynamics can be derived \cite{Bianchi:2010gc}
in the limit of a large number $N$ of polyhedral faces. Here the deformation dynamics of 
the polyhedron is a secondary contribution after the configuration entropy of the polyhedron, 
which can be readily developed from the statistical mechanics of polymers. 

The dynamics of  the elementary polyhedron, the tetrahedron, can be exactly solved and 
semi-classically  quantized through the Bohr-Sommerfeld procedure \cite{Bianchi:2011ub, 
Bianchi:2012wb}.  The volume spectrum arising from quantizing this classical system has 
shown remarkable agreement with fully quantum calculations.  If the tetrahedron is the 
``hydrogen atom'' of space, the next complex polyhedron, the pentahedron ($N=5$), can
be considered as the analogue of the helium atom. Just as the full range of atomic physics
phenomena occurs first in the helium atoms with its two electrons, the pentahedron 
represents the first space configuration that puts the ideas discussed above to a non-trivial 
test. 

The dynamical system corresponding to the isochoric pentahedron with fixed face areas 
has a four-dimensional phase space compared with two dimensional phase space of the 
tetrahedron. The integrability of a Hamiltonian system in a phase space with more than two
dimensions is not assured. Non-integrable Hamiltonian systems exhibit a wide variety of 
interesting behaviors, including Hamiltonian chaos. The incompressibility of Hamiltonian 
flows leads to a very specific form of chaos \cite{Gutzwiller:1990,Licht:1992,Almedia:1988}.
A well known example is the three-body Kepler problem, which is non-integrable, in contrast
to the two-body problem \cite{Poincare:1892} and exhibits chaos at the classical level. In the
analogous quantum system, the helium atom, the chaotic dynamics exhibits itself through
the mixing of bound and continuum states.

There are two distinct classes of polyhedra with five faces, the triangular prism and a
pyramid with a quadrilateral base. The latter forms a measure zero subset of allowed
configurations as its construction requires reducing one of the edges of the triangular
prism to zero length. This process imposes an algebraic constraint between configuration
variables making this a co-dimension 1 configuration. 

In this article we first briefly review the symplectic Kapovich-Millson phase space of
polyhedral configurations in Section \ref{sec:poly-and-phase}. In Section
\ref{sec:poly-recon} we outline a method by which it is possible to uniquely construct a
triangular prism (or quadrilateral pyramid) for each point in the four-dimensional phase
space and review a method due to Laserre \cite{Laserre} for computing the volume of any
polyhedron from its face areas and their normals. We introduce a convenient labeling
convention of the faces of the pentahedron in Section \ref{sec:poly-config}. With these
tools in hand we are then able to compute the trajectory of a pentahedron starting from a
given point in phase space with the volume of the pentahedron serving as its
Hamiltonian. This volume preserving (isochoric) evolution keeps the areas of the faces of
the pentahedron fixed and only allows for changes in their shape and orientation. In this
way the edge lengths and vertex positions of the pentahedron are free to vary, in effect
the pentahedron is smoothly deformed, keeping its volume constant. In Section
\ref{sec:stability} we examine the local dynamical stability of configurations of the
pentahedron throughout the phase space and map the distribution of the local Lyapunov
exponents as function of the volume of the pentahedron. To explore the stability further
we compute the time evolution of intermediate time Lyapunov exponents which suggest the
presence of chaos at small volumes.  We conclude our article with a summary of our results
and a brief analysis of their possible implications in Section \ref{sec:conclusion}.

\section{Polyhedra and Phase Space}
\label{sec:poly-and-phase}


A convex polyhedron is a collection of faces bounded with any number of vertices.
A theorem by Minkowski \cite{Minkowski:1897} states that the areas $A_l$ and normals 
$\vec{n}_l$ of each face are sufficient to uniquely characterize a polyhedron. If we define
$\vec{A_l} = A_l \vec{n}_l$ then the polyhedral closure relationship
\begin{equation}
  \label{eqn:mink}
  \sum_{l} \vec{A}_l = 0,
\end{equation}
is a sufficient condition on $\vec{A}_{l}$ to uniquely define a polyhedron with $N$ faces. 
The space of shapes of polyhedra $\mathcal{P}_N$ with $N$ faces was further investigated 
by Kapovich and Millson \cite{Kapovich:1996}. The shape space is defined as the space
of all polyhedra modulo to their orientation in three-dimensional space:
\begin{equation}
  \label{eqn:kap-mil-def}
  \mathcal{P}_N = \left\{ \vec{A}_l \mid \sum_l \vec{A}_l = 0, | \vec{A}_l | = A_l \right\} / SO(3).
\end{equation}
The shape space of convex polyhedra with $N$ faces is thus $2(N-3)$-dimensional;
in particular, the shape space of the tetrahedron ($N=4$) is two-dimensional and that of
the pentahedron ($N=5$) is four-dimensional.
As Kapovich and Millson showed, this space admits a symplectic structure, which can be
defined by introducing a Poisson bracket for any two functions $f(\vec{A}_l), g(\vec{A}_l)$ as
\begin{equation}
  \label{eqn:poisson-brackets}
  \{ f, g \} = \sum_l \vec{A}_l \left( \frac{\partial f}{\partial \vec{A}_l} \times \frac{\partial g }{\partial \vec{A}_l} \right).
\end{equation}

Canonical variables with respect to this Poisson bracket are defined on $\mathcal{P}_N$ as
follows. One first defines the vector sum $\vec{p}_k = \sum_{l=1}^{k+1} \vec{A}_l$ of the
first $k+1$ oriented faces. The ordering/labeling of these normals is not important for
the physical shape of the polyhedron. Minkowski's theorem ensures that any set of vectors
which obey \eqref{eqn:mink} will produce a \emph{unique} polyhedron. 

The canonical momenta in the KM space are defined as $p_k = | \vec{p}_k |$ and the conjugate
positions are given by the angle between $\vec{p}_k \times \vec{A}_{k+1}$ and $\vec{p}_k
\times \vec{A}_{k+2}$. Using \eqref{eqn:poisson-brackets} one can verify that $\{q_k,
p_k'\} = \delta_{kk'}$. We refer the reader to \cite{Kapovich:1996, Bianchi:2010gc} for detailed explanations and
proofs.

\begin{figure*}[ht]
  \centering
  \includegraphics[width=0.4\textwidth,clip, trim=0.05cm 0.05cm 0.05cm 0.5cm]{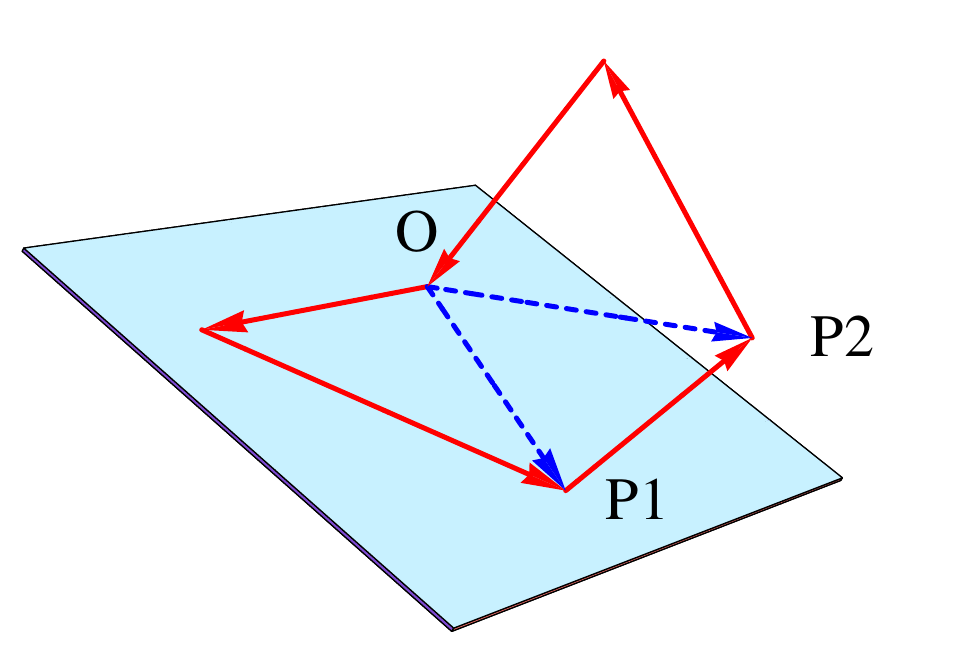}
  \includegraphics[width=0.3\textwidth, clip, trim=0.5cm 0.5cm 0.5cm 0.25cm]{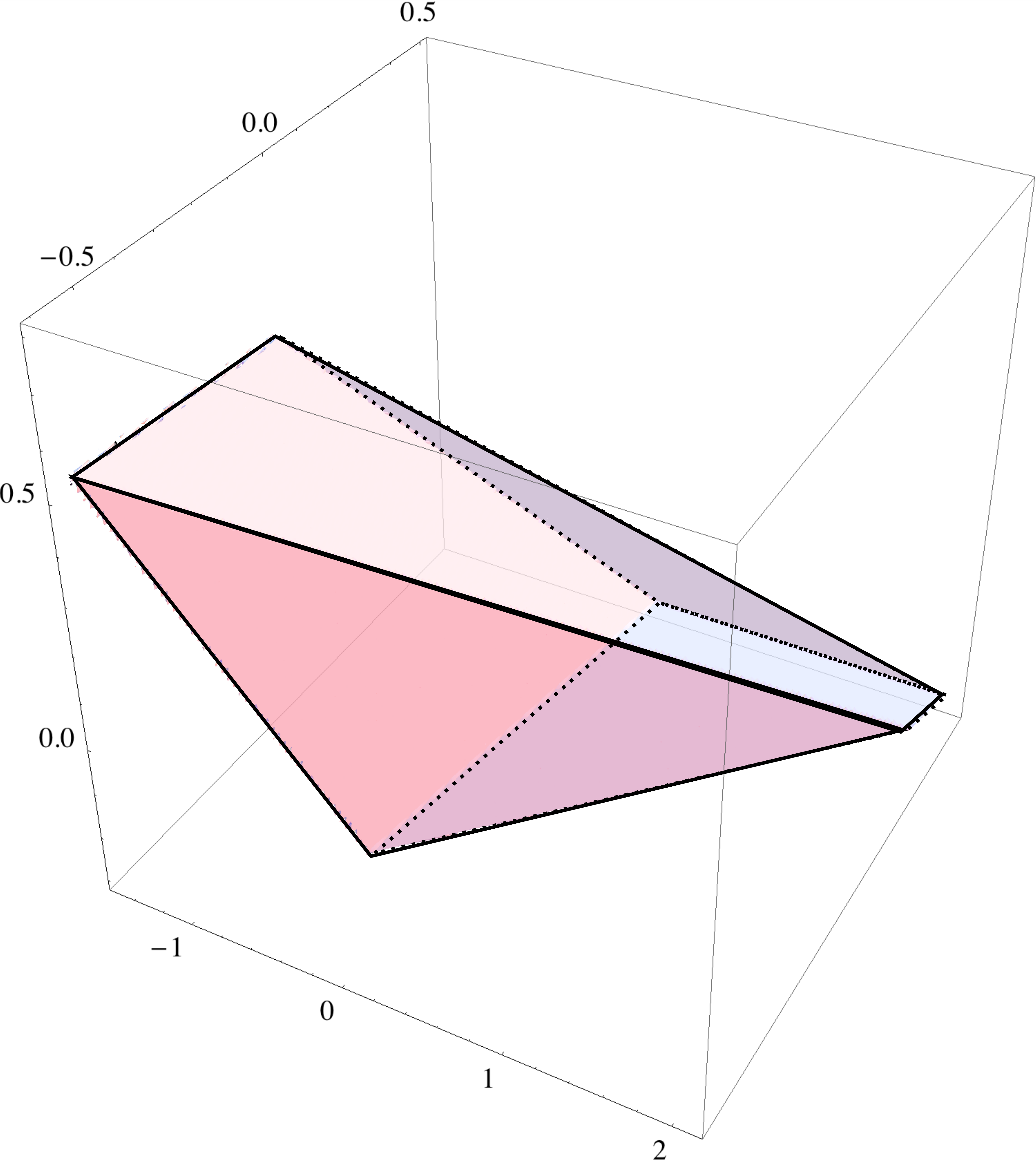}
  \caption{An example configuration of the system in the polygon representation (left),
    the phase space coordinates plotted here are $z=\{0.3, 0.4, 0.9, 0.91\}$. The normal
    vectors are plotted as the red solid arrows and the momentum vectors are plotted as
    the dashed blue arrows. We also show a rendering of the associated polyhedron
    (right). All polyhedral faces have area fixed to one, so all polygonal edges have unit
    length .}
  \label{fig:polygon-simple}
\end{figure*}

These quantities may be visualized by representing the polyhedron as a polygon with edges
given by the vectors $v_{i} = \sum_{l=1}^{i}{\vec{A}_l}$. Note that this generally yields a 
non-planar polygon.  Now consider systematically triangulating this polygon starting from
one vertex, which is chosen as the origin. The inserted edges are the conjugate momenta $p$ 
and the angles between each of these edges are the conjugate positions. An illustration of the
pentagon associated with a pentahedron in shown in \figref{fig:polygon-simple}.

The momentum vectors $\vp_k$ define axes about which subsets 
of the normals rotate as the polyhedron is deformed; the angles $q_k$ measure the rotation 
about these axes.  The collection of coordinates $(p_k,q_k)$ thus completely describes the
 ``bending flow'' of the polyhedron.

\begin{figure}[tb]
  \centering
  \includegraphics[width=0.4\textwidth]{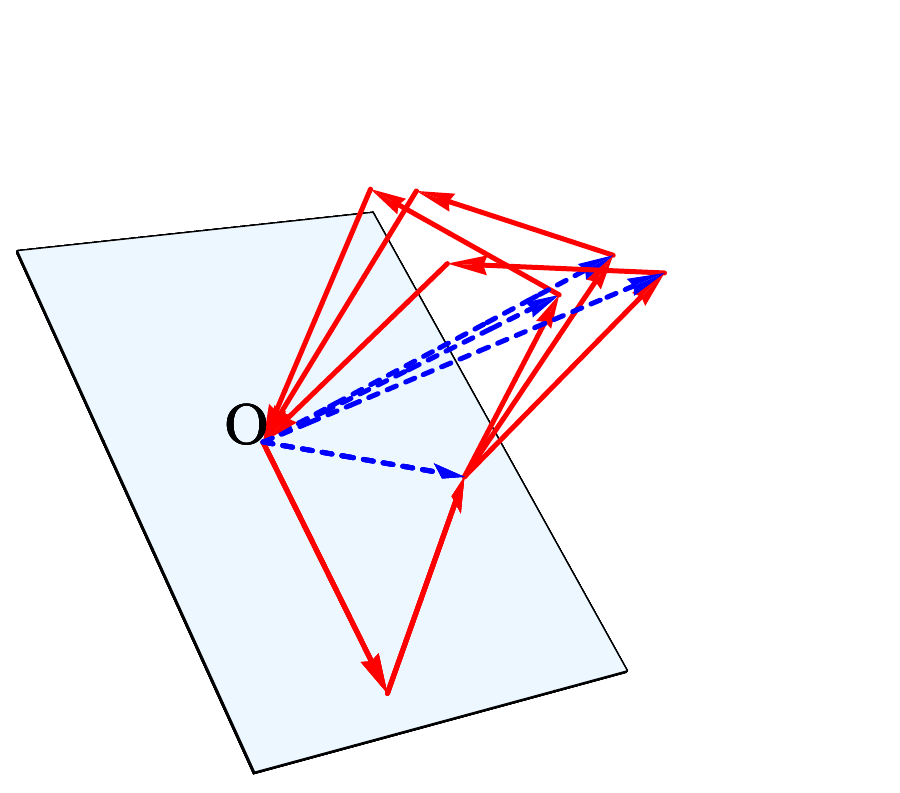}
  \includegraphics[width=0.3\textwidth]{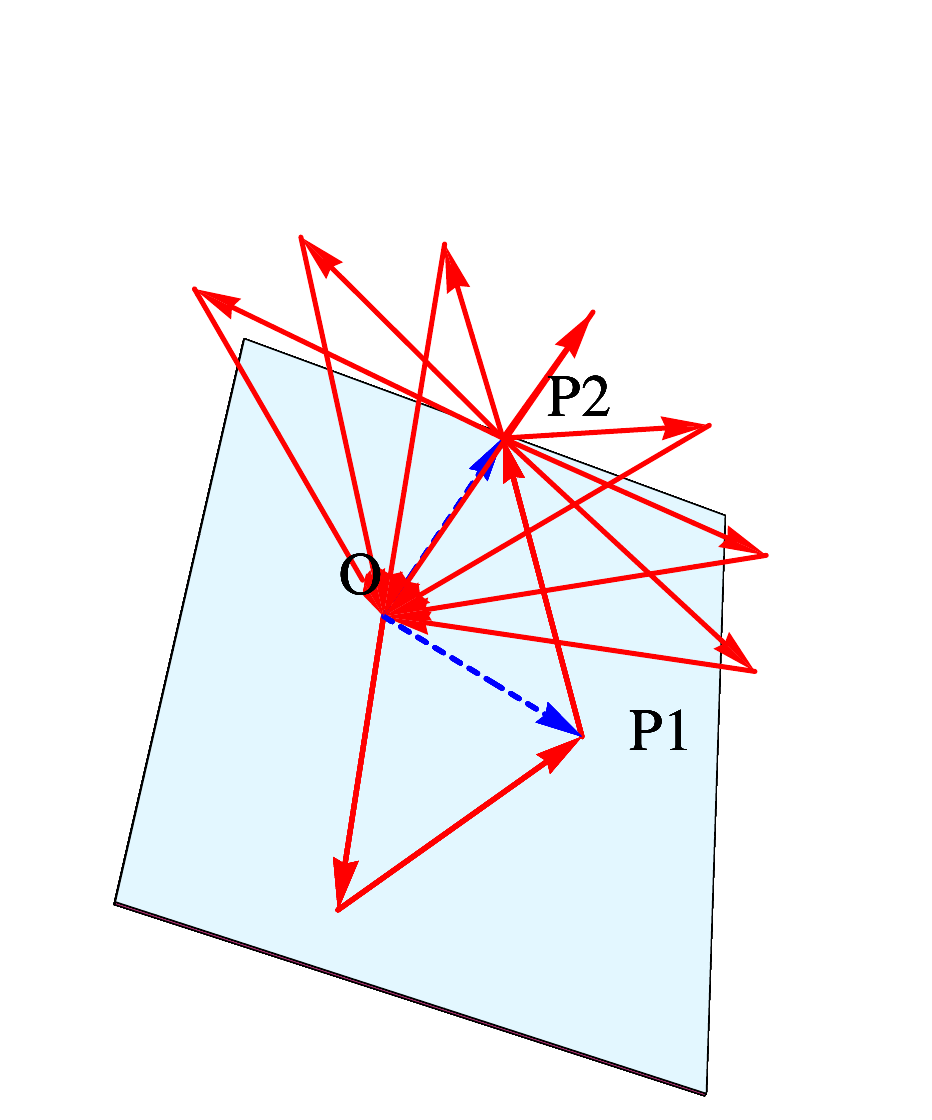}
  \caption{\label{fig:polygon-q-rotate} A superposition of configurations showing the
    rotation of $q_1$ (left) and $q_2$ (right) through $\pi/2$ and $\pi$ radians. The
    remaining coordinates are held fixed. The normal vectors are plotted as the red solid
    arrows and the momentum vectors are plotted as the dashed blue arrows. All faces have
    area fixed to one. }
  \vspace{-0.5cm}
\end{figure}

The momentum vectors $\vec{p}_1, \vec{p}_2, \ldots, \vec{p}_{N}$ partition the polygon
into a set of coupled subsystems. In \figref{fig:polygon-q-rotate} we show several pentagons 
illustrating the variation of the angles $q_1$ (left) and $q_2$ (right), while holding the 
remaining phase space variables fixed. 

It is not trivial to calculate the volume of a general polyhedron directly from the phase
space coordinates. Haggard has computed an analytic formulation for the
\cite{Haggard:pc}. Alternatively it is straightforward, if cumbersome, to compute the
volume of a polyhedron from its normals and face areas.  We seek a mapping $F:z \to n(z)$
between a point $z$ in the phase space and the normals $n(z)$ associated with that point.

We now specialize our discussion to the pentahedron, although it is easily generalized
to higher polyhedra.  One of the normals can be completely defined in terms of the four
others by the closure of the pentahedron. Since we require that the normal vectors be 
normalized, so that the areas of the faces are fixed, this leaves two degrees of freedom 
per vector. After enforcing closure we thus need seven equations to completely specify 
the pentahedron in space. The definitions of the canonical variables provide four equations, 
leaving three free quantities. We fix these three final components by picking an orientation 
of the coordinate system. This is sufficient to specify $F:z \to n(z)$.

It is convenient to orient the polyhedron in such a way that one of the faces lies in the 
$x$--$y$ plane and its normal is oriented in the negative $z$ direction $(0,0,-1)$. 
To fix the final coordinate direction we require that one of the edges of this face is 
parallel to the $x$ direction. This fixing the normal of the adjacent face to have
components $(0, n_{y}, n_{z})$. We are free to label our normals in any order. Let $n_5$
be the normal fixed by closure, $n_1$ be the normal in the $x$--$y$ plane and $n_2$ the
normal to the face which has the edge parallel to $x$. 

It is useful to explicitly derive the inverse mapping $F^{-1}:n(z) \to z$,
i.~e.\ the map between a pentahedron represented by a set of normals and face areas and 
the set of canonical variables $z=\{q_1,q_2,p_1,p_2\}$. The canonical variables can be 
obtained from the normals and areas as:
\begin{align}
  \label{eqn-canonical-tri-prism}
  p_1 &= | \vec{A}_1 + \vec{A}_2 | = | -\vec{A}_5 - \vec{A}_4 - \vec{A}_3 |, \notag\\
  p_2 &= | \vec{A}_1 + \vec{A}_2 + \vec{A}_3 | = | -\vec{A}_5 - \vec{A}_4 |, \notag\\
  q_1 &= \mbox{angle}\left\{ \vec{p}_1 \times \vec{A}_2 ,\vec{p}_1 \times \vec{A}_3 \right\}, \notag\\
  q_2 &= \mbox{angle}\left\{ \vec{p}_2 \times \vec{A}_3 ,\vec{p}_2 \times \vec{A}_4 \right\}.
\end{align}
\begin{widetext}
It is helpful to introduce the following vectors 
\begin{align}
  \label{eqn-tri-prism-pvecs}
  \vp_0 &= \va_1, \notag \\
  \vp_1 &= \va_1 + \va_2 = \vp_0 + \va_2, \notag \\
  \vp_2 &= \va_1 + \va_2 + \va_3 = \vp_1 + \va_3, \notag \\
  \vp_3 &= \va_1 + \va_2 + \va_3 + \va_4 = \vp_2 + \va_4,
\end{align}
the canonical angles can be written as 
\begin{align}
  \label{eqn-tri-prism-canonical-angles}
  q_1 &= \mbox{angle}\left\{ \vp_0 \times \vp_1, \vp_1  \times \vp_2 \right\}, \notag\\
  q_2 &= \mbox{angle}\left\{ \vp_1 \times \vp_2, \vp_2  \times \vp_3 \right\}.
\end{align}
\end{widetext}

To ensure that the angles $q_1, q_2$ are uniquely defined over the phase space we take
care to compute the signed angles between the vectors $\vp_0 \times \vp_1$ and $\vp_1
\times \vp_2$. We define the signed angle between
two vectors $\vec{a}, \vec{b}$ with a reference vector $\vec{r}$  in \eqref{eqn:signed-angle}. In
practice the reference vector is the appropriate momentum vector $\vp_k$. This is
sufficient to define $F^{-1}$.
\begin{align}
  \label{eqn:signed-angle}
  S &= \frac{ \mid \vec{a} \times \vec{b} \mid } { \mid \vec{a} \mid \mid \vec{b} \mid}, \notag\\
  C &= \frac{ \vec{a} \cdot \vec{b} } { \mid \vec{a} \mid \mid \vec{b} \mid}, \notag\\
  \alpha &= \vec{r}\cdot\left(\vec{a}\times\vec{b} \right), \notag\\
  \theta &= \begin{cases} 
    \arctan(C, S) \quad \mbox{if } \sgn \alpha > 0, \\
    2\pi - \arctan(C, S) \quad \mbox{if } \sgn \alpha < 0
    \end{cases}
 \end{align}
Given the choice of orientation of our coordinate system along with face areas
$A_{1,\ldots, 5}$, the polyhedral configuration is uniquely determined by the following
five vectors
\begin{align}
  \label{eqn-norms-tri-prism}
  \vn_1 &= (0,0,-1), \notag\\
  \vn_2 &= (0,n_{2y},n_{2z}),\notag\\
  \vn_3 &= (n_{3x},n_{3y},n_{3z}),\notag\\
  \vn_4 &= (n_{4x},n_{4y},n_{4z}),\notag\\
  \vn_5 &= -\frac{1}{A_5} \left(A_1 \vn_1 + A_2 \vn_2 + A_3 \vn_3 + A_4 \vn_4 \right).
\end{align}
We use normalization to fix the magnitudes of $n_{2y}, n_{3x} n_{4x}$, leaving 5
unknown components. The following set of equations arise from inserting \eqref{eqn-norms-tri-prism} into the definitions of
$|\vp_1|,|\vp_2|$ and  $|\vp_2| = |\vp_1 + \va_2|$,
\begin{align}
  p^2_1  &= A_1^2 + A_2^2 + 2A_1 A_2 \vn_1 \cdot \vn_2,   \label{eqn-tri-p1-square}\\
  p^2_2  &= A_5^2 + A_4^2 + 2A_4 A_5 \vn_4 \cdot \vn_5,   \label{eqn-tri-p2-square}\\
  p^2_2  &= p^2_1 + A_3^2 + 2A_2 A_3 \vn_2 \cdot \vn_3 + 2A_1 A_3 \vn_1 \cdot \vn_3,   \label{eqn-tri-p1p2-square} 
\end{align}
To close the system we use the definitions of the canonical angles
\eqref{eqn-tri-prism-canonical-angles}. At this point the system can be inverted
numerically to give solutions for the components of the normals in terms of the phase
space variables. Particular care needs to be taken to ensure that unique configurations of
normals are obtained for each set of phase space variables $z$. Further care needs to be
taken in picking a sign for the components whose magnitudes are fixed by
normalization. The signs of these components vary across the space in a non trivial
fashion. We introduced a set of heuristics to suggest which set of signs might be
appropriate, the inverse map $F^{-1}$ is vital for testing and rejecting trial solutions.

It is helpful to note that the set $\{\va_1, \va_2, \va_3, -\vp_2\}$ can be taken as the
faces of a tetrahedron with canonical momentum $\vp_1$ and conjugate angle $q_1$. The
dynamics of this tetrahedron are independent of the second bending angle $q_2$. The set of
equations to be inverted can be factored into a set of 3 equations for $n_{2z}, n_{3y},
n_{3z}$ in $p_1, p_2, q_1$ and a set of 2 equations for $n_{4y}, n_{4z}$ which are dependent upon the full set
of canonical variables. The former can be solved analytically. Fixing all face areas to
$A$ and restricting $q_1$ to the first quadrant gives
\begin{widetext}
\begin{align}
  \label{eqn-norms-tri-prism-tet}
  n_{2z} &= 1 - \frac{p_1^2}{2A^2},\\
  n_{3y} &= \frac{ -(4A^2-p_1^2)(A^2+p_1^2 - p_2^2) \pm |\cos(q_1)|  \sqrt{(-4A^2 + p_1^2)\left(A^4 + (p_1^2-p_2^2)^2 - 2 A^2 (p_1^2 + p_2^2)\right)} } 
    {4A^2 p_1 \sqrt{4A^2 - p_1^2}} ,\\
  n_{3z} &= \frac{1}{4A^2 p_1}\left(p_1(A^2+ p_1^2-p_2^2) \pm |\cos(q_1)| \sqrt{ (-4A^2+p_1^2)(A^4 + (p_1^2 - p_2^2)^2 - 2A^2 (p_1^2 + p_2^2))}   \right).
\end{align}
\end{widetext}
The full solution with different face areas and with $q_1$ taking values in all quadrants is obtained
using a computer algebra system. A Newton based root finder \cite{Norcedal:2006} is
then used to find simultaneous solutions to $q_2 = \mbox{angle}\left\{ \vp_1 \times
  \vp_2, \vp_2 \times \vp_3 \right\}$ and  \eqref{eqn-tri-p2-square}. 

The numerical integration of the dynamical system, which is based around gradients of the
Hamiltonian, requires considerable numerical accuracy in $F$.  The choice of initial guess
for the root search when solving for $n_{4y}, n_{4z}$ is vital for obtaining a solution. To improve this process an
empirical PDF with density inversely proportional to the euclidean norm of the residual
vector from the root finding routine was generated by evaluating trial solutions over a
grid in the space. This PDF was simultaneously refined while being used to
generate initial guesses for the roots.

The construction method outlined above can be generalized to develop a chain of
polynomials for $N>5$ allowing at least numerical evaluation of $F$.

\subsection{The shape of the phase space}
\begin{figure}[ht]
  \centering
  \includegraphics[width=0.3\textwidth]{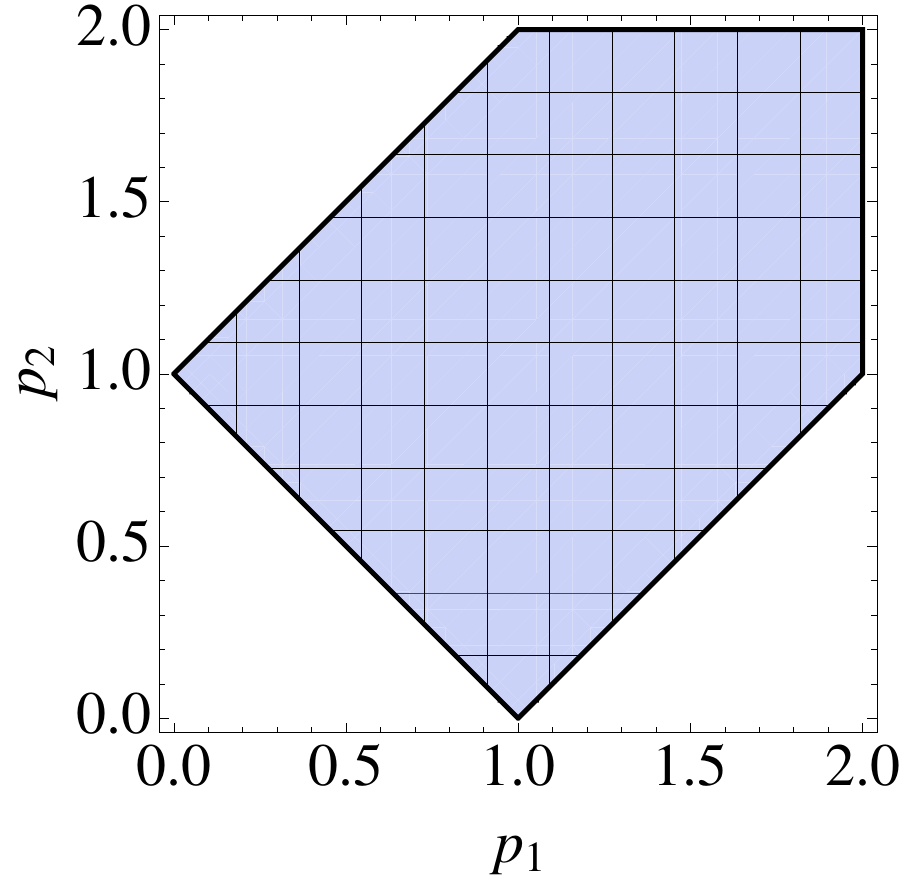}
  \caption{The shaded area shows the allowed region  of momentum space, satisfying
    \eqref{eqn:momentum-ranges} and \eqref{eqn:momentum-ranges-joint},
  for a system with $A_1 = A_2 = A_3 = A_4 = A_5 = 1$}
\label{fig:momentum-allowed}
\end{figure}

The geometric structure of the polyhedron itself, particularly the fixed face areas,
induces certain restrictions upon the phase space. The position space is $2\pi$ periodic
by construction. The momentum space is restricted by the areas of the faces, from the triangle inequality 
\begin{align*}
p_1 &= |\vec{A}_1 + \vec{A}_2| \le |\vec{A}_1| + |\vec{A}_2|, \\
p_2 &= |\vec{p}_1 + \vec{A}_3| \le |\vec{p}_1| + |\vec{A}_3|.
\end{align*}
Heron's formula for the area of a triangle can be used to simplify the above inequalities, 
\begin{equation}
  \label{eqn:heron}
  \mathfrak{A}(a,b,c) = \frac{1}{4}\sqrt{(a+b+c)(a+b-c)(a-b+c)(b+c-a)}
\end{equation}
where $a,b,c$ are the edges of the triangle. Consider the triangles $\Delta_1 =
\{\vec{A}_1, \vec{A}_2, \vec{p}_1\}$, $\Delta_2 = \{\vec{p}_1, \vec{A}_3, \vec{p}_2\}$ and
$\Delta_3 = \{\vec{p}_2, \vec{A}_4, \vec{A}_5\}$. For the system to be in a reasonable
configuration we require that the area of each of these triangles be non zero, i.e that the terms under the radical in \eqref{eqn:heron} be positive. This restriction on $\Delta_1$ and $\Delta_3$ implies
\begin{align}
  \label{eqn:momentum-ranges}
  |A_1 - A_2 | &\le p_1 \le |A_1 + A_2 |, \notag \\
  |A_4 - A_5 | &\le p_2 \le |A_4 + A_5 |. 
\end{align}
Applying this to $\Delta_2$ gives
\begin{align}
  \label{eqn:momentum-ranges-joint}
  A_3 + p_1 > p_2, \notag \\
  A_3 + p_2 > p_1, \notag \\
  p_1 + p_2 > A_3, \notag \\
  A_3 + p_1 + p_2 > 0.
\end{align}
The analysis in the remainder of this article is focused on the case where all the areas are fixed equal to one, the allowed momentum space for this case is shown in \figref{fig:momentum-allowed}.

\section{Hamiltonian \& Polyhedral Reconstruction}
\label{sec:poly-recon}

Following from \cite{Bianchi:2012wb, Bianchi:2011ub, Bianchi:2010gc} we use the volume of the pentahedron at a given point in the phase space as the Hamiltonian. This ensures that trajectories generated  by Hamilton's equations will deform the pentahedron while maintaining a constant volume. Unlike the case of the tetrahedron, it is not trivial to compute the volume of the general pentahedron in terms of the canonical coordinates. Although elegant direct geometric expressions for the volume of any polyhedron exist \cite{Haggard:pc},  these do not give particularly tractable expressions in terms of the phase space variables. 

Consider a vector field $\vec{F}(x) = \frac{1}{3}\vec{x}$, using the divergence theorem we can find the volume of a polyhedron
\begin{equation}
  \label{eqn:vol-div}
 V = \int_{\Omega} \vec{\nabla} \cdot \vec{F} d \Omega = \oint_s \vec{F} \cdot \vec{n} ds = \sum \frac{1}{3} \vec{x_i} \cdot \vec{n}_i
\end{equation}
where $\Omega$ is the interior of the polyhedron, $\vec{x}_i$ is a point on the i'th face and $\vn_i$ is the normal to that face. 

We can compute the volume of a polyhedron specified as a set of normals and areas using \eqref{eqn:vol-div} once we know the location of a point upon each face. To obtain this we need the edges, or equivalently vertices, of the polyhedron. These can be found by \emph{reconstructing} the polyhedron from its normals and areas, a process originally due to Laserre \cite{Laserre} and more recently outlined in \cite{Bianchi:2010gc}. This Laserre reconstruction procedure works for any $N$ and requires only the minimization of a quadratic function, in some cases algebraic results can be directly obtained.

The reconstruction process works by pushing a set of $N$ infinite planes, some set of distances $\vec{h} =\{h_1, \ldots, h_N\}$ from the origin along the normals of the system. The areas of the polyhedral faces formed by the union of the half-spaces defined by these planes can be computed as a function of the heights $A(\vec{h})$. A numerical minimization routine \cite{Norcedal:2006} is then used to extract the set of heights which minimize the form
\[
 \mid A(\vec{h}) - \mathcal{A} \mid^2 
\]
where $\mathcal{A}$ is the set of desired areas. (In passing we note that by carrying out this procedure after flipping the sign of the normals we can obtain the chiral dual of a given polyhedron). The set of vectors $\{ h_1 \vec{n}_1, \ldots, h_{N} \vec{n}_{N}\}$ can then be used in \eqref{eqn:vol-div} to compute the volume of the system.

\begin{figure*}
  \centering
  \includegraphics[width=0.29\textwidth]{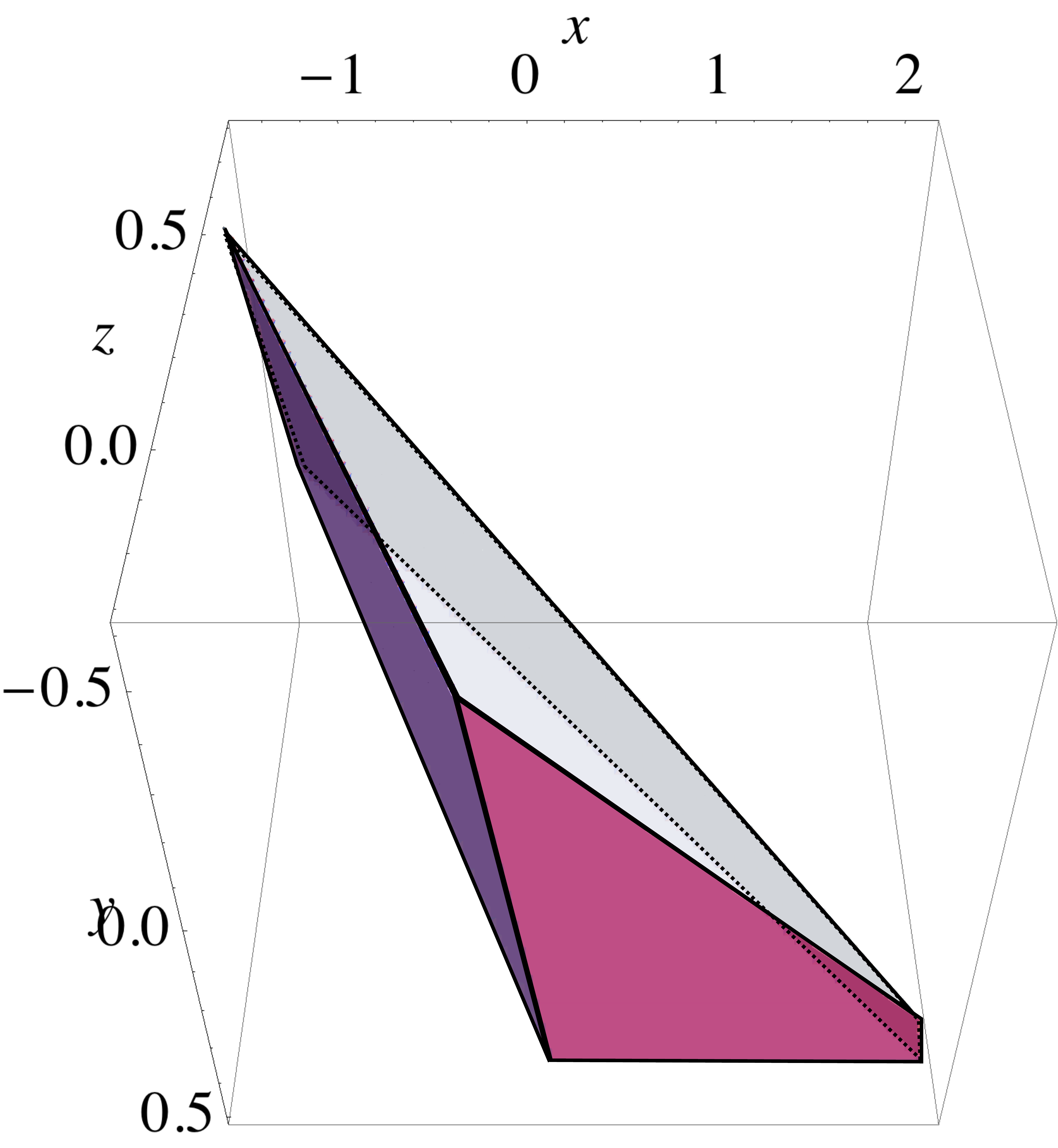}
  \hspace{0.1cm}
  \includegraphics[width=0.29\textwidth]{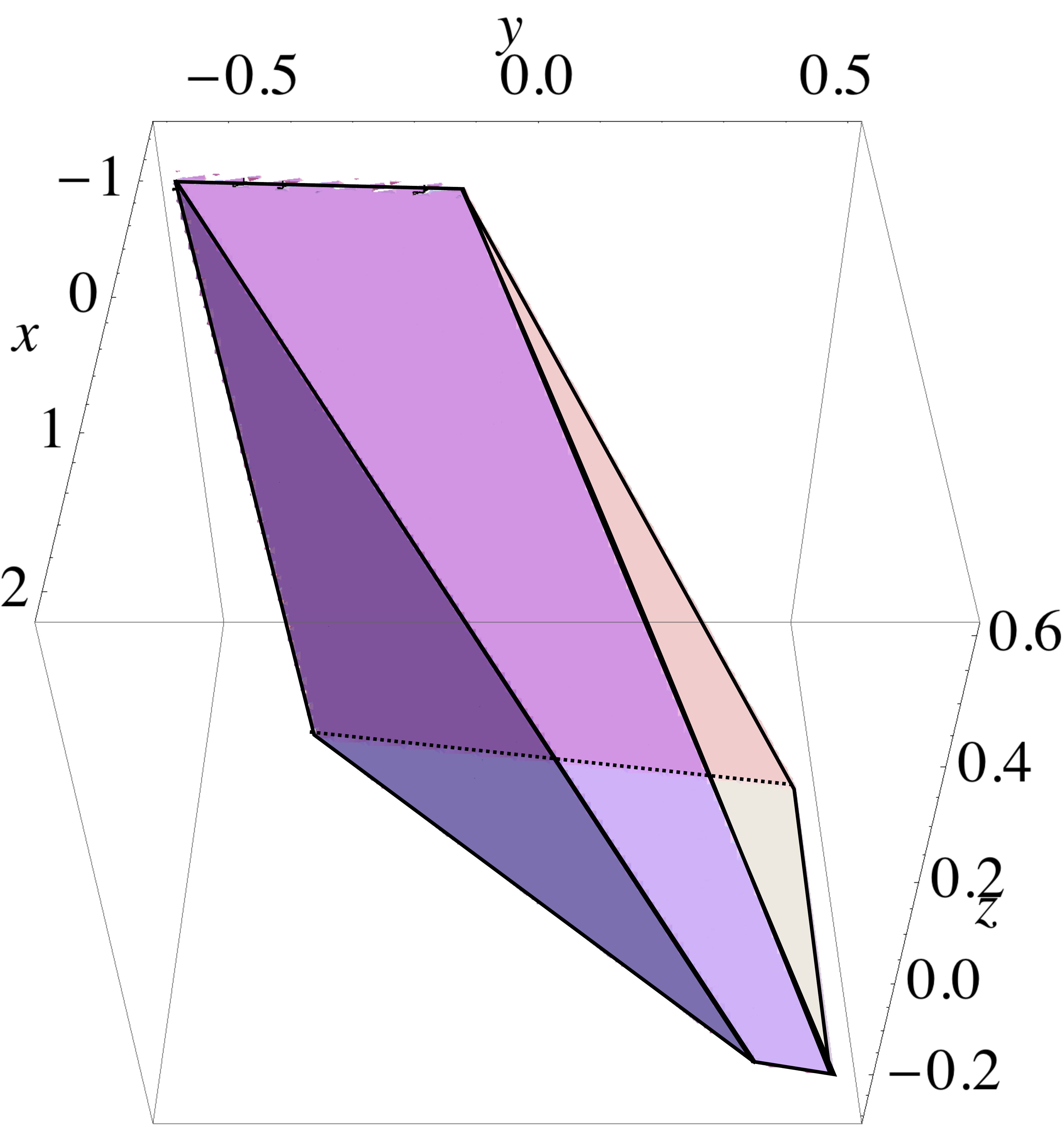}
  \hspace{0.1cm}
  \includegraphics[width=0.29\textwidth]{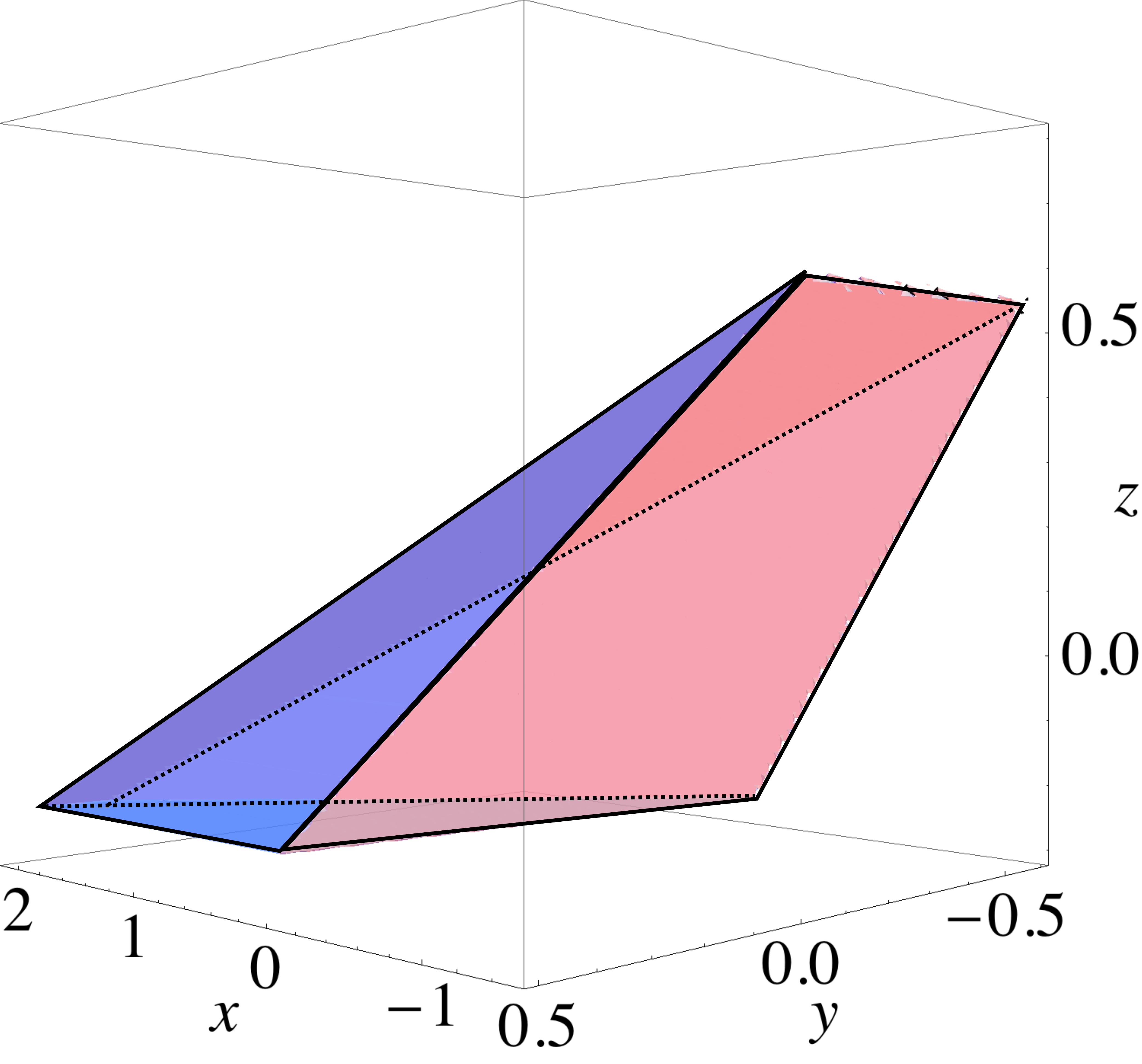}
  \caption{A set of renderings of the reconstructed pentahedron corresponding to the point $z_1 = \{0.3, 0.4, 0.9, 0.91\}$ in the KM space, all face areas set are to 1. Where visible internal edges have been drawn with dashed lines, external edges with solid lines.}
  \label{fig:polyhedron-simple}.
\end{figure*}

As an example we reconstruct the pentahedron corresponding to $z_1 = \{0.3, 0.4, 0.9, 0.91\}$ shown in \figref{fig:polygon-simple} with all the face areas fixed to 1. Applying our mapping to compute the normals $F(z_1)$ gives:
\begin{align}
  \label{eqn:normals-specific}
  \vec{n}_1 &= \{0,0,-1\}, \notag \\
  \vec{n}_2 &= \{0,0.803,0.595\}, \notag \\
  \vec{n}_3 &=\{0.249,-0.114,0.961\}, \notag \\
  \vec{n}_4 &= \{0.301,-0.921,0.244\}, \notag \\
  \vec{n}_5 &= \{-0.550,0.232,-0.801\}. 
\end{align}
After applying the reconstruction routine we obtain the following set of heights
\begin{equation}
  \vec{h} = \{0.237, 0.239,  0.236, 0.239, 0.237\}
\end{equation}
with a corresponding volume of $0.396621$. The heights found are of a similar magnitude because the face areas are all fixed to the same value, this precludes very anisotropic configurations. A set of renderings of the polyhedron is shown in \figref{fig:polyhedron-simple}.

In the remainder of this article we will consider the case where all faces of the
polyhedron have the unit area. We can estimate some limiting values of the volume in this
case. A regular equilateral triangular prism should be close to the maximum volume. If the
triangular edges have length $\ell$ and  the vertical quadrilateral edges have length $h$, 
the volume is $V_{\rm prism} = A \ell $, where $A$ denotes the common area of all faces. 
Using \eqref{eqn:heron} we obtain the triangular edge length $\ell = \frac{2\sqrt{A}}{3^{1/4}}$, 
and the quadrilateral edge length $h = A/\ell$ so $V_{\rm prism} = \frac{3^{1/4}}{2} A^{3/2}
\approx 0.658 A^{3/2}$. 
The volume of a regular square pyramid can also be obtained from $V_{\rm pyr} = \frac{1}{3} A h$ 
where $A$ is the area of the base and $h$ is the height of the pyramid. Again using 
\eqref{eqn:heron} we obtain $h = \sqrt{15A}/2$ and so $V_{\rm pyr} = \frac{\sqrt{15}}{6}A^{3/2}
\approx 0.645 A^{3/2}$. 
It is interesting to note that the regular square pyramid has a slightly smaller volume
than the regular triangular prism.

The volume is bounded from below by zero, in the limit of collinear normals the volume is
certain to vanish.  A general feature of this Minkowski polyhedral reconstruction
process is that if the area of one face is smoothly shrunk to zero the system will
smoothly deform into a polyhedron of a lower order \cite{Bianchi:2010gc}. 

We can estimate the behavior of the volume as it approaches zero by considering the
extreme cases of \eqref{eqn:momentum-ranges-joint}. As we shrink the second momentum $p_2$
to zero then $p_1 \to A$. In this limit the area of the triangle $\Delta_3 = \{A_4, A_5,
p_2\}$ goes to zero, we suppose that the volume of the system is then entirely dominated
by the set of vectors $\{A_1, A_2, A_3, p_2\}$, the volume of the associated tetrahedron \cite{Bianchi:2011ub,Bianchi:2012wb}
is
\begin{equation}
  \label{eqn:tet-vol}
  V_{tet}^2 = \frac{8}{9}\frac{\mathfrak{A}(A_1, A_2, p_1) \mathfrak{A}(p_1, p_2, A_3)}{p_1} \sin(q_1),
\end{equation}
for a fixed value of $p_1$ on the boundary of the allowed space the volume goes smoothly
to zero as $p_2$ is taken to its extreme value. Taking a unit area system for simplicity
with $q_1 = \pi/2, p_1 = 1, p_2 = 1 - x$ and expanding \eqref{eqn:tet-vol} for small $x$
we obtain
\[
\left.V_{tet}\right|_{A=1} = \frac{1}{\sqrt{6}}-\frac{x}{3 \sqrt{6}}-\frac{x^2}{3 \sqrt{6}} + \mathcal{O}(x^3)
\]
The same construction can be carried out \emph{mutatis mutandis} to obtain a limiting a form of
the volume as the first momentum is shrunk to zero. The derivatives of the volume along
the limiting direction are well behaved in this limit.

\begin{figure}[t]
  \centering
  \includegraphics[width=0.25\textwidth]{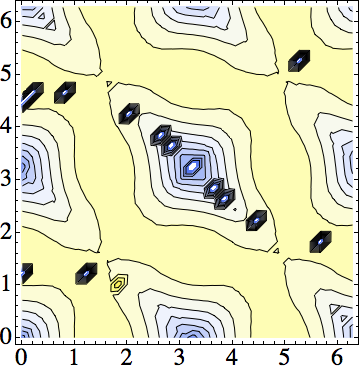}
  \caption{A section in the $q_{1}, q_{2}$ plane through the Hamiltonian evaluated at $p_{1} = p_2 = 0.94$, all face areas are fixed to 1. The contours are isochors, the color scheme is brighter at larger volumes.}
  \label{fig:ham-single-qq}
\end{figure}

\begin{figure}[b]
  \centering
  \includegraphics[width=0.3\textwidth]{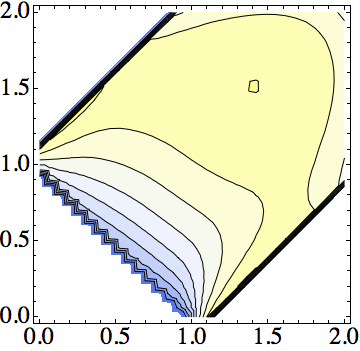}
  \caption{A section in the $p_{1}, p_{2}$ plane through the Hamiltonian evaluated at $q_{1} = 0.3, q_{2} = 0.4$, all face areas are fixed to 1. The contours are isochors , the color scheme is brighter at larger volumes.}
  \label{fig:ham-single-pp}
\end{figure}

A section of the polyhedral volume through the $q_1,q_2$ plane is shown in
\figref{fig:ham-single-qq}, the spots are regions where the numerical reconstruction
routine failed. The contours show lines of constant volume (isochors), cooler colored regions
correspond to smaller volumes. The volume is visibly $2\pi$ periodic in both position
variables. The structure of the Hamiltonian in this projection suggests that of
a complicated coupled harmonic oscillator in a periodic space.  In
\figref{fig:ham-single-pp} we show a projection of the Hamiltonian in the $p_1, p_2$ plane
at fixed $q_1, q_2$.  The lowest volume regions in this plot are found where the momenta
take their minimum allowed values. The dependence of the volume upon the momenta is not
trivial, the Hamiltonian appears to be non separable in its current form.

\begin{figure*}[ht]
  \centering
  \includegraphics[width=0.5\textwidth]{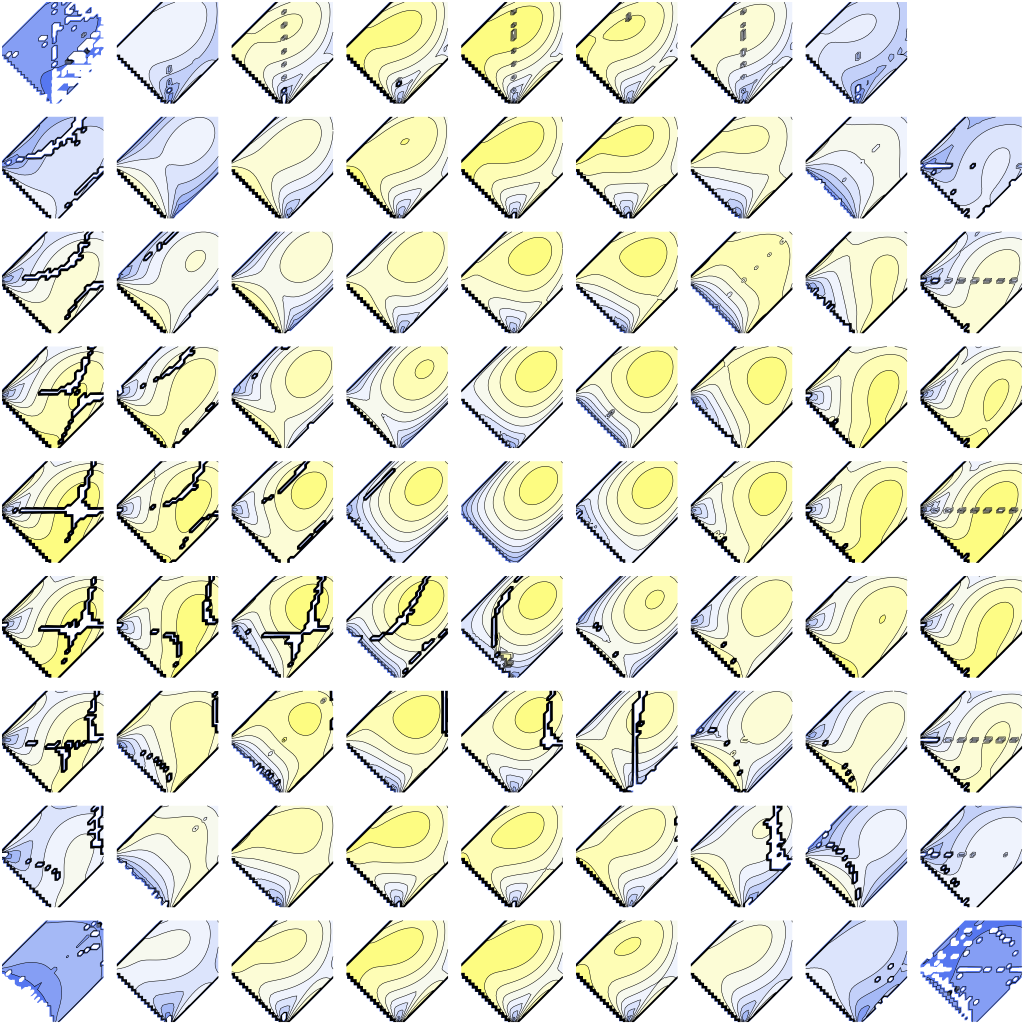}
  \caption{Sections through the Hamiltonian in the $p_1, p_2$ space. The plots span a range of zero to $\pi$ in $q_1, q_2$ with $q_2$ increasing along the vertical axis and $q_1$ along the horizontal axis. Cooler blues correspond to smaller volumes, brighter yellows correspond to larger volumes.} 
  \label{fig:ham-grid}
\end{figure*}

In \figref{fig:ham-grid} we show a set of projections in the $p_1, p_2$ plane which
together span the angle space from $\pi$ to $2\pi$. The behavior in the other quadrants of
the $q_1, q_2$ range is very similar to the one shown here, we present this reduced set of
plots to give the reader some intuition of the variation in the volume over the whole
space.  The scarring visible in some of the subplots is a result of the failure of the
root finding heuristic along boundaries where $n_{4x}$ flips sign. The volume is symmetric
about the line $q_1 = q_2$ and the maximum volume is modulated sinusoidally as the angles
pass through a period. The maximum volume is seen to be on the $p_1 = p_2$ line for $q_1 =
q_2 \simeq 3 \pi /2 $, in the other angular quadrants the maximal volume is obtained at multiples of $\pi/2$.
\begin{widetext}
The plots around maximum volume are symmetric about the line $p_1 = p_2$, here the $p_1, p_2$ projection
is well described by the sum of the areas of the triangles $\Delta_1, \Delta_2,
\Delta_3$ as given by \eqref{eqn:heron}
\begin{equation}
  \label{eqn:ham-ana}
H(q_1, q_2, p_1, p_2) \propto  \mathfrak{A}(p_1, A_1, A_2)\frac{\sin(q_1)}{\sqrt{p_1}} + \mathfrak{A}(p_1, p_2, A_3)\frac{\sin(q_1)\sin(q_2)}{(p_1 p_2)^{1/4}} + \mathfrak{A}(p_2, A_4, A_5)\frac{\sin(q_2)}{\sqrt{p_2}},
\end{equation}
this is a good description across the whole range of momenta and for small angular deviations around $\pi/2$.
\end{widetext}
This choice of the form of \eqref{eqn:ham-ana} is motivated by the expression for the
squared volume of a tetrahedron obtained in \cite{Bianchi:2011ub}.  The radicals in the
denominator are needed to fix the dimensions. The canonical momenta have dimensions of
area so the Hamiltonian must have dimensions of $A^{3/2}$.  The full Hamiltonian is
deformed in a more complex way when moving away from multiples of $\pi/2$ in the angle
space.  The approximate Hamiltonian \eqref{eqn:ham-ana} may give insight into the dynamics
of the system in the large volume limit, we shall present an analysis of the trajectories
of this and the full Hamiltonian in a future work.

\begin{figure*}[ht]
  \centering
  \includegraphics[width=0.5\textwidth]{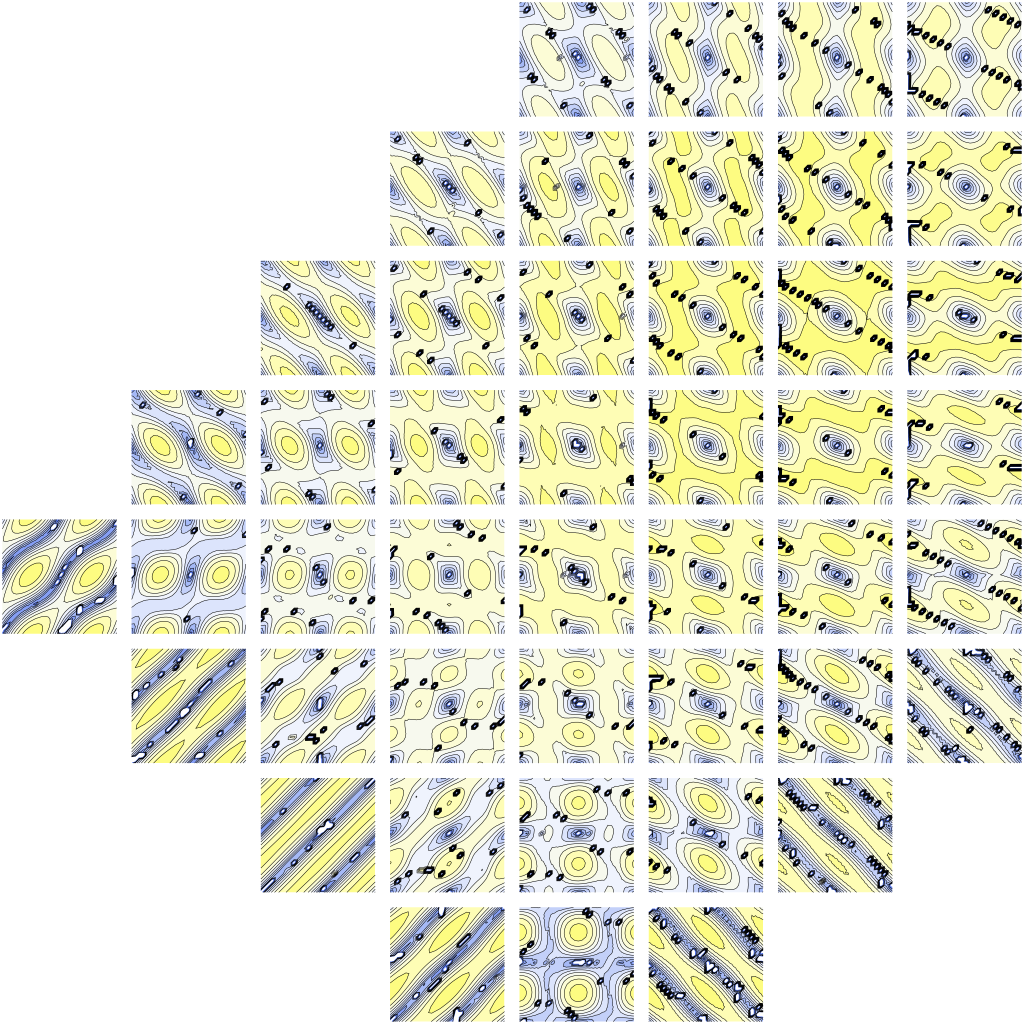}
  \caption{Sections through the Hamiltonian in the $q_1, q_2$ space. The plots span the range $[0,2]$ in $p_1, p_2$ with $p_2$ increasing along the vertical axis and $p_1$ along the horizontal axis. Cooler blues correspond to smaller volumes, brighter yellows correspond to larger volumes.} 
  \label{fig:ham-grid-qq}
\end{figure*}

An extended set of projections in the $q_1,q_2$ plane is shown in
\figref{fig:ham-grid-qq}, the set of plots shown spans the full range of both momentum
variables. The periodic structure shown in \figref{fig:ham-single-qq} exists throughout
the whole space, however the periodicity is carried by volume maxima at small momenta
(bottom left) and volume minima at large momenta (top right). 

The volume of the triangular prism has a very complex structure over the allowed regions
of the KM phase space. The volume is observed to be smooth and differentiable across the
space. Numerical examination of the volume in the limit that the system approaches the
momentum boundaries of the phase space indicate a smooth approach to zero.

\section{Configurations}
\label{sec:poly-config}

Due to our particular choice of the orientation of the $\mathbb{R}^3$ coordinate axes we
are able to uniquely distinguish the normals in our implementation of the triangular
prism.  Our numerical map from a phase space point $z$ to the normals $n(z)$ makes
identification of each of the remaining normals clear and gives them a natural ordering.

We adopt a similar scheme to Haggard \cite{Haggard:pc} and label the possible configurations of the
pentahedron in terms of which normals correspond to square or triangular faces. Note that
a full reconstruction of the pentahedron is required to determine the number of edges (or
order) associated with each face and therefore with each normal. Fixing all the faces to
have the same area we are no longer able to uniquely identify a particular square or
triangular face. However we are still able to distinguish the order of the faces.

We label the distinct configurations by concatenating the indices of the triangular
faces. A configuration where the third and fourth faces are triangular would be labeled
$34$ or equivalently $43$. We label the configurations in lexical order to remove this
ambiguity. The set of available configurations $\mathcal{C}$ is
\[
\mathcal{C} = \{12, 13, 14, 15, 23, 24, 25, 34, 35, 45 \}.
\]

\begin{figure}[ht]
  \centering
  \includegraphics[width=0.3\textwidth]{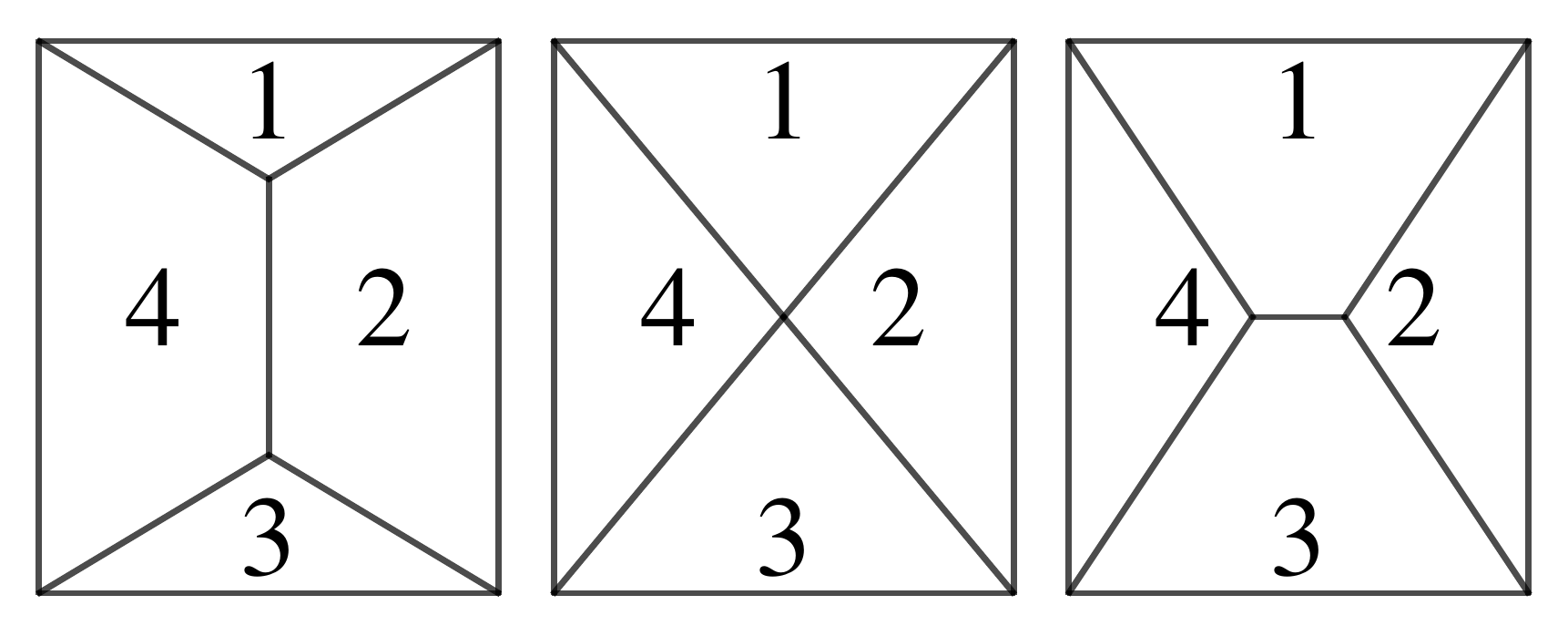}
  \caption{Schlegel diagrams showing the transformation of a $13$ configuration to a $24$ configuration. The $5$ face, not shown in these figures, forms the back face of the graphs.}
  \label{fig:config-transform}
\end{figure}

The triangular prism can be smoothly deformed from one configuration to several
others. These deformations involve a two step process.  First the smooth shrinking of one
the edges shared by two square faces to zero, creating a quadrilateral pyramid in the
limit that the edge vanishes. The vertex at the apex of this new pyramid is split into an
edge which now borders two newly square faces. An example of this process in shown in
\figref{fig:config-transform}, here a $13$ triangular prism configuration is deformed into
a $24$.

\begin{figure}[htb]
  \centering
  \includegraphics[width=0.3\textwidth]{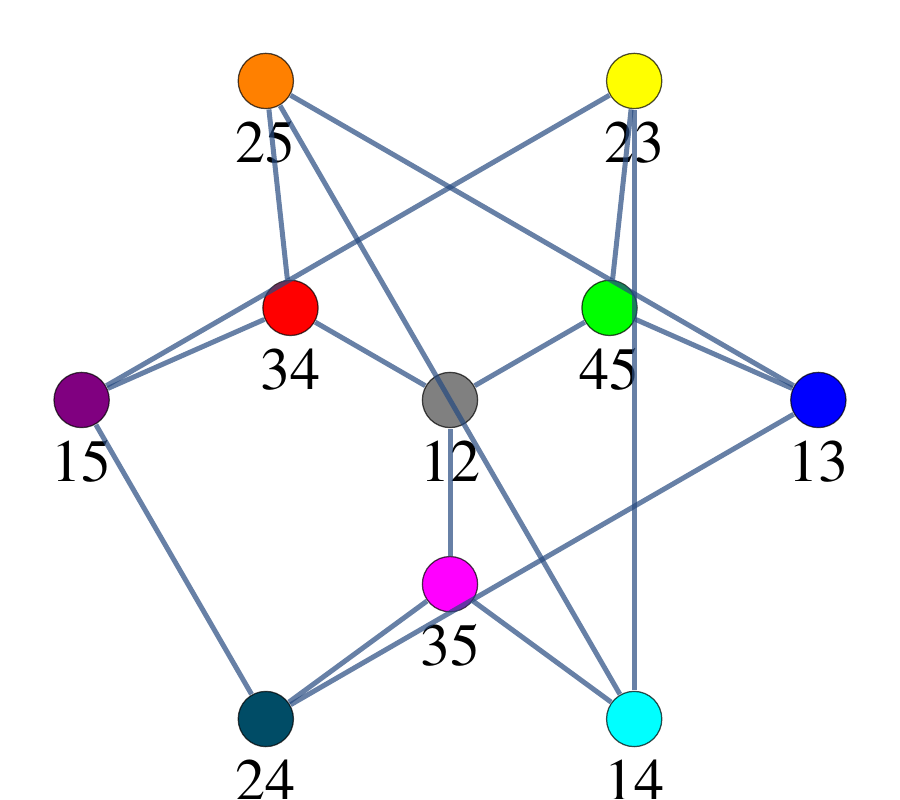}
  \caption{Allowed transitions between configurations. The vertex colors correspond to the colors plotted in the configuration scan figures \figref{fig:config-1}, \figref{fig:config-2}}
  \label{fig:config-graph}
\end{figure}

\begin{figure*}
  \centering
  \includegraphics[width=0.5\textwidth]{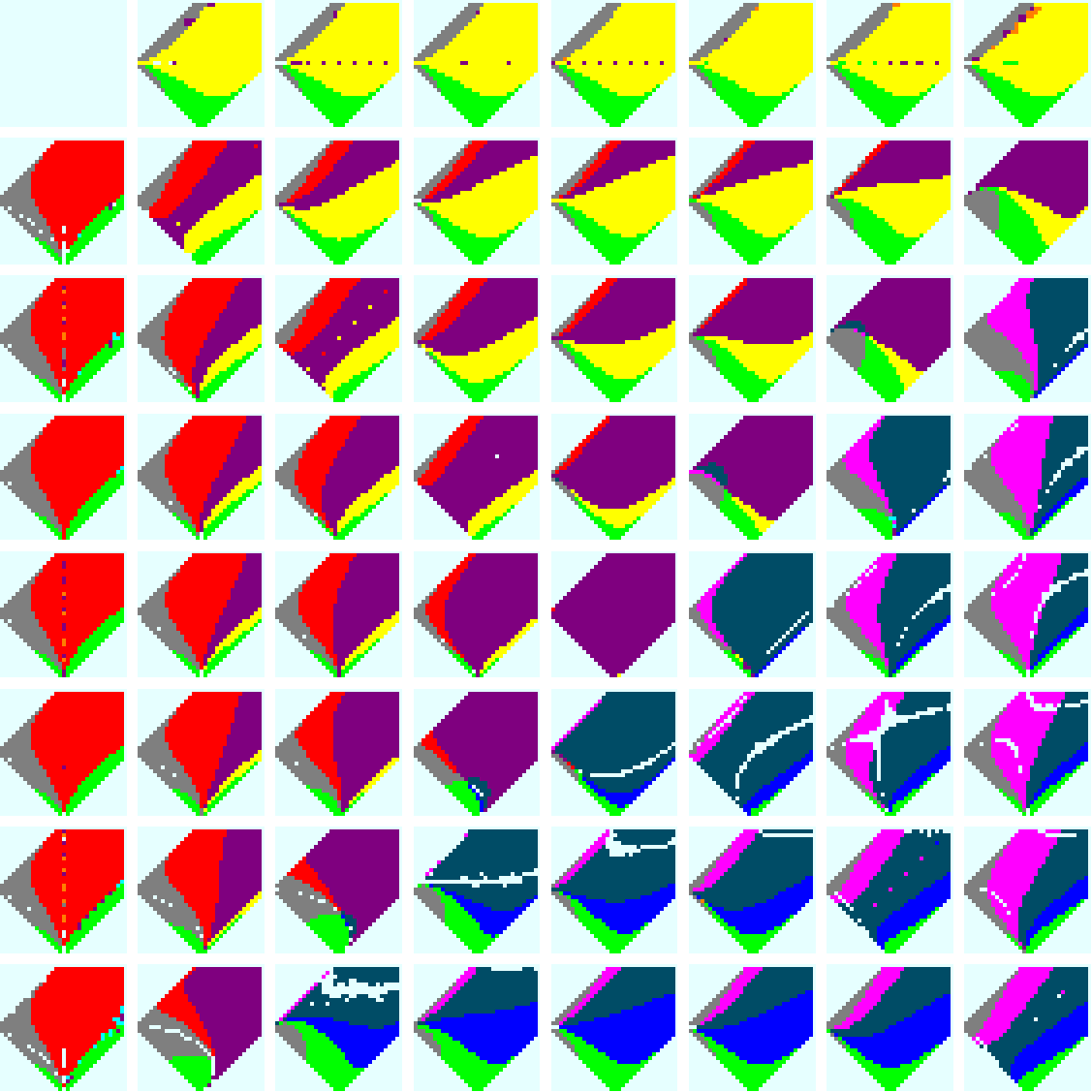}
  \caption{Slices through the phase space in the $p_1,p_2$ plane over $q_1,q_2$ in $[0,\pi]$ showing the different configurations of the triangular prism. The projection here corresponds to that shown in \figref{fig:unstab-grid}.}
  \label{fig:config-1}
\end{figure*}

\begin{figure*}
  \centering
  \includegraphics[width=0.5\textwidth]{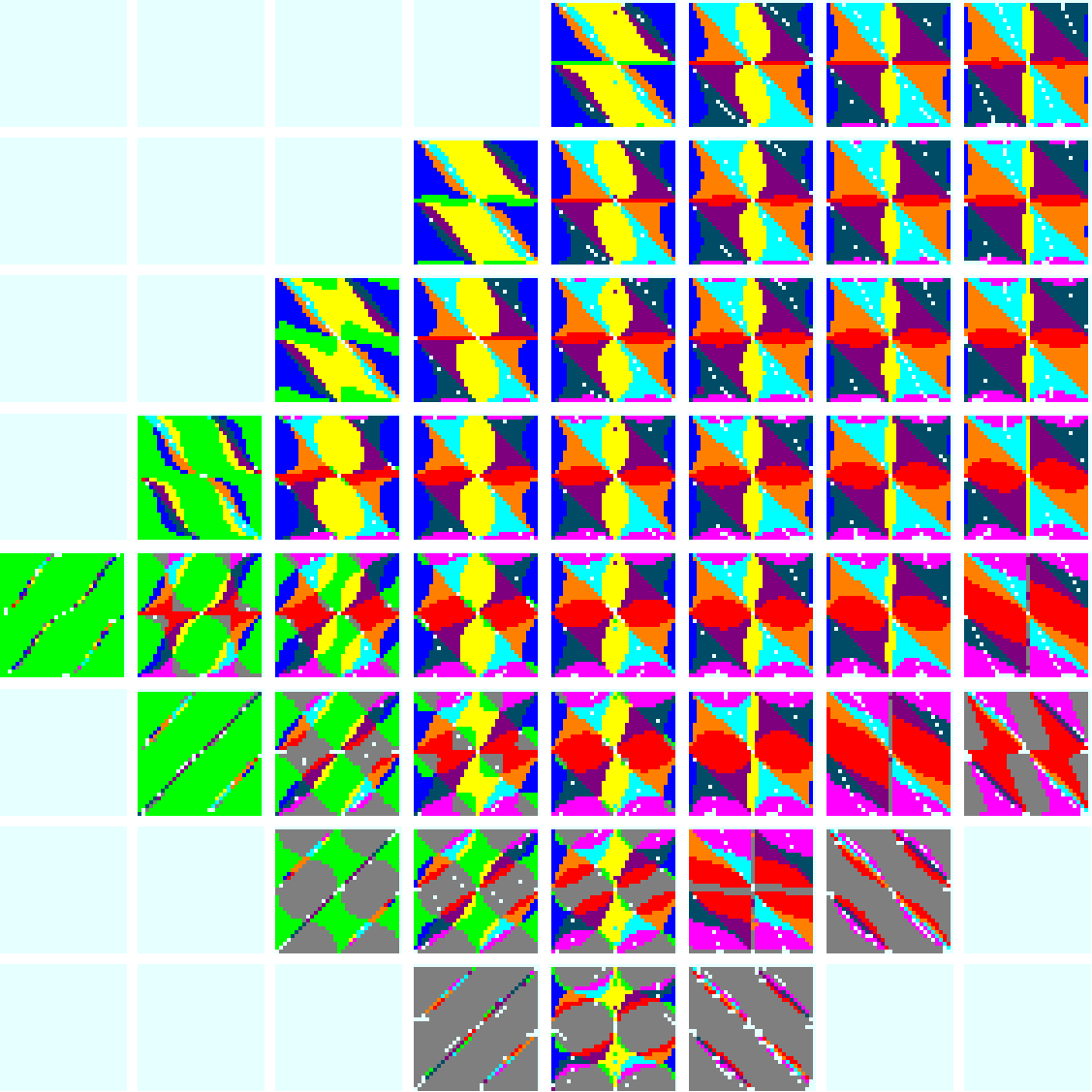}
  \caption{Slices through the phase space in the $q_1,q_2$ plane over the full range of $p_1,p_2$, showing the different configurations of the triangular prism. The projection here corresponds to that shown in \figref{fig:unstab-grid-qq}.}
  \label{fig:config-2}
\end{figure*}

The undirected graph in \figref{fig:config-graph} shows the set of allowed
deformations. For example the transformation $13 \to 24$ shown in \figref{fig:config-transform} 
would be represented as moving from the blue vertex to the dark green vertex.  The colors of the
vertices in the graph serves as a key to the colors shown in \figref{fig:config-1} and
\figref{fig:config-2}. These figures show the configuration of the system across the phase
space in the $p_1, p_2$ and $q_1, q_2$ planes. The graph can be used to understand which
boundaries in these figures can be smoothly crossed and which are disallowed.

The set of momentum-space projections shown in \figref{fig:config-1} show a complex but
smooth distribution of the interfaces between configurations through the space. Many of
the boundaries represent allowed transitions. Several disallowed transitions are also
observed. The $34$ configuration (red) is spatially adjacent to the $45$ configurations
(light green) in the first column of figures, in the first row the $23$ configuration
(yellow) is adjacent to the $12$ (gray). These unphysical transitions may serve as a kind
of separatrix in the momentum space. The position-space projections shown in
\figref{fig:config-2} exhibit a greater complexity in the boundaries between the
configurations. However disallowed transitions are not obvious in this projection.

\section{Stability}
\label{sec:stability}

The Lyapunov exponents of a dynamical system characterize the rate of separation of  pairs 
of initially adjacent trajectories \cite{Licht:1992, Gutzwiller:1990, Almedia:1988}. 
Positive Lyapunov exponents provide a universal signature of Hamiltonian chaos. 
Their computation requires a reliable algorithm for the integration of Hamilton's equations,
which respects their symplectic nature. Unfortunately, integrating Hamilton's equations 
for the isochoric pentahedron 
has proven to be rather challenging, because the Hamiltonian is not analytically known
in terms of the phase space variables. All derivatives of the Hamiltonian with respect
to the phase space variables must be computed numerically. This introduces numerical 
errors into the Hamiltonian equations of motion, which may not be relevant over short
segments of the trajectory, but can introduce large errors in long trajectories, which are
needed for the determination of the global Lyapunov exponents. To avoid this complication, 
we here pursue a simpler method which gives useful insights into the integrability and 
stability of the pentahedron as a dynamical system.

The Kolmogorov-Sinai (KS) entropy gives the rate at which a dynamical system destroys 
information (creates entropy); it is given by the sum over all positive global Lyapunov 
exponents of the system. Lacking a reliable numerical method for calculating the GLE's, 
we will estimate the KS entropy in terms of the local Lyapunov exponents. In doing so, 
it is worth noting that local instability does not guarantee the global instability of a trajectory. 
For example, the well known $x^2y^2$ system is almost everywhere locally unstable, 
but it admits nontrivial periodic trajectories that are globally stable. In general, 
the phase space averaged LLE's will provide an upper bound for the KS entropy, 
because the local divergence between two trajectories can be balanced by local 
convergence in other places, leading to a smaller global rate of divergence along 
the entire trajectory. On the other hand, while a small number of globally stable 
trajectories cannot be excluded in a dynamical system that is locally unstable in 
a large fraction of phase space, most trajectories will be globally unstable.

To calculate the local Lyapunov exponents (LLE) of a Hamiltonian system we consider 
the small local deviation of a pair of trajectories $\delta z(t)$, which can be computed 
by linearizing Hamilton's equations:
\begin{equation}
  \label{eqn-lle-ham-eqn}
  \delta \dot{z}(t) = \mathcal{H}(t,z) \delta z(t),
\end{equation}
where $\mathcal{H}$ is the Jacobian matrix of the Hamiltonian evaluated at $z$
\begin{equation}
  \label{eqn-lle-matrix}
  \mathcal{H}(t,z) = 
  \begin{pmatrix}
    -\partial_{pq} H & -\partial_q^{2} H \\
    \partial_{p}^2 H & \partial_{pq} H .
  \end{pmatrix}
\end{equation}
Solving \eqref{eqn-lle-ham-eqn} gives $\delta{z}(t) = \exp(\lambda_z t) u$ where
$\lambda_z$ is an eigenvalue of $\mathcal{H}(z)$ and $u$ is the associated
eigenvector. Eigenvalues with positive (negative) real components are associated with
exponentially diverging (converging) trajectories, purely imaginary eigenvalues represent
a periodic motion. Although the LLE's of a system typically do not correspond to the true
or global Lyapunov exponents (GLE's) they serve as a local measure of the departure of a
pair of trajectories and give a good indication as to the stability of the system at a
particular point. We define a point to be stable if all the positive real components of
the eigenvalues of $\mathcal{H}$ are zero, i.e if $\max\{\Re \lambda_{z}^{+} \} = 0$. Our
numerical explorations of the phase space have shown the Hamiltonian to be relatively
smooth and continuous, computing the local stability over a finite grid should give a
reasonable approximation to the overall stability of the system.

If the entire phase space was found to be unstable it would be reasonable to conclude that
the system is globally unstable and that the associated GLE's would be positive. However
what can we conclude if we find a finite set of stable points amongst some set of unstable
points? 

\begin{figure*}[t]
  \centering
  \includegraphics[width=0.5\textwidth]{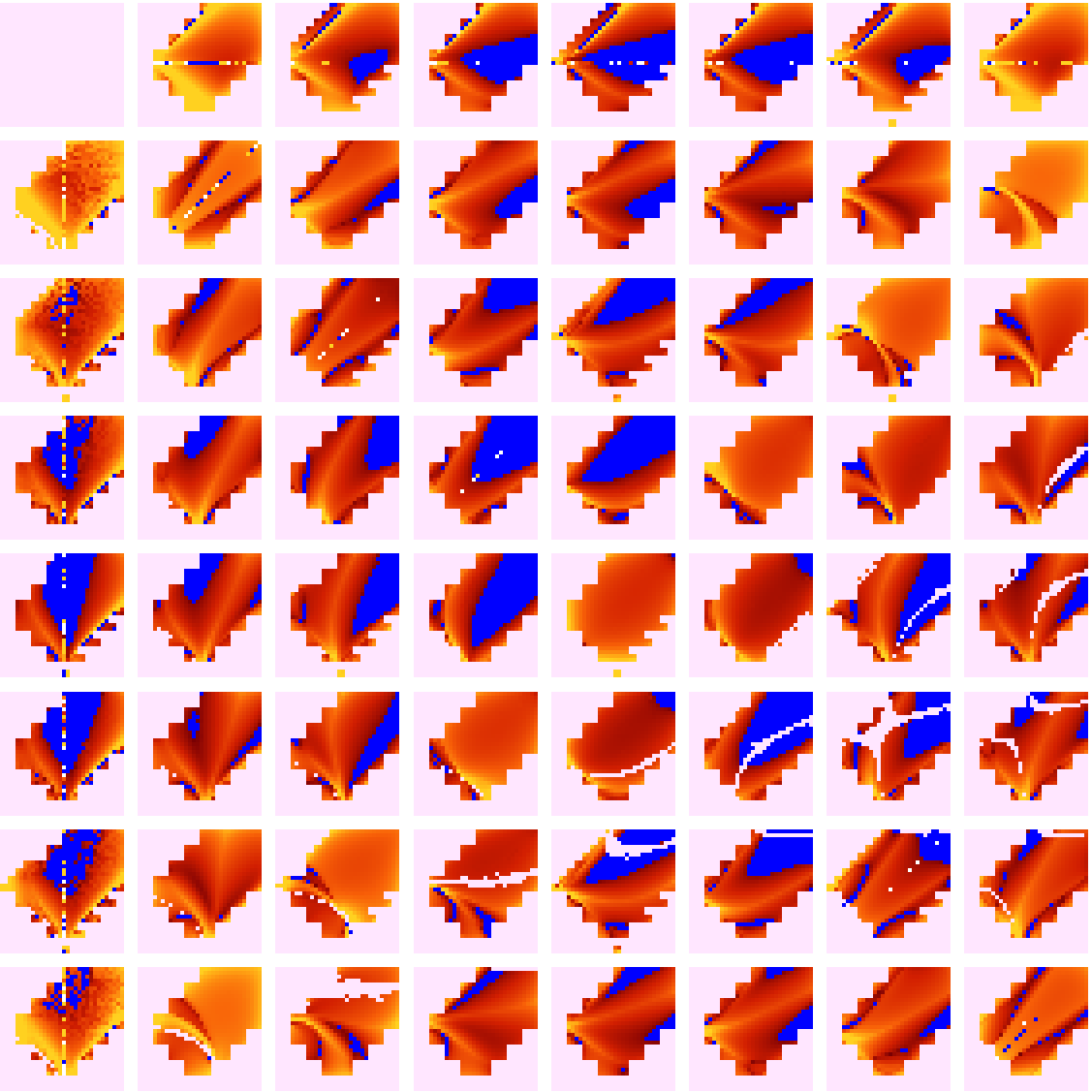}
  \caption{The local Lyapunov exponents computed for sections in the $p_1, p_2$ over $q_1,q_2$ in $[0,\pi]$. The royal blue regions are stable and hot colors (red, orange, yellow) represent progressively more unstable regions. The code failed to converge in the light cyan regions. Note that here only LLE's in the range $\lambda \in [0, 1]$ are plotted here.}
  \label{fig:unstab-grid}
\end{figure*}

\begin{figure*}[b]
  \centering
  \includegraphics[width=0.5\textwidth]{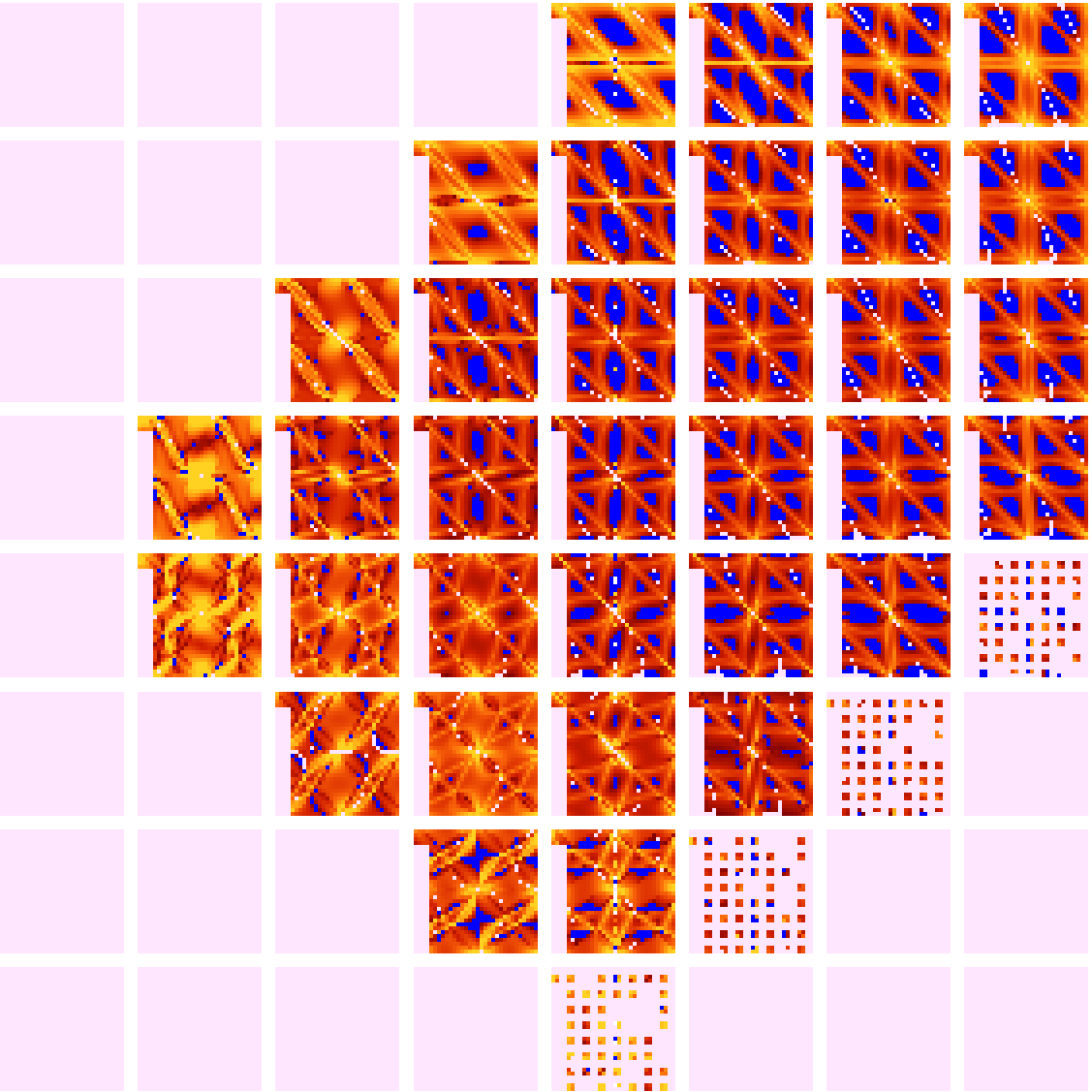}
  \caption{The local Lyapunov exponents computed for sections in the $q_1, q_2$ over the full $p_1,p_2$ range. The royal blue regions are stable and hot colors (red, orange, yellow) represent progressively more unstable regions. The code failed to converge in the light cyan regions. Note that here only LLE's in the range $\lambda \in [0, 1]$ are plotted here.}
  \label{fig:unstab-grid-qq}
\end{figure*}

We compute the local stability over a grid of $2^{20}$ points in the phase space, partial
projections in the $p_1,p_2$ and $q_1,q_2$ planes showing the largest positive eigenvalues
are shown in \figref{fig:unstab-grid} and \figref{fig:unstab-grid-qq}. These figures show
that most of the space is locally unstable (hot colors), however sizable stable regions
(dark blue) do appear. The large values (light yellow) in the stability distributions
shown are clearly correlated with the boundaries in the configuration space (see
\figref{fig:config-1} and \figref{fig:config-2}). Regions close to disallowed transitions
appear very unstable. 

\begin{figure}[b]
  \centering
  \includegraphics[width=0.4\textwidth]{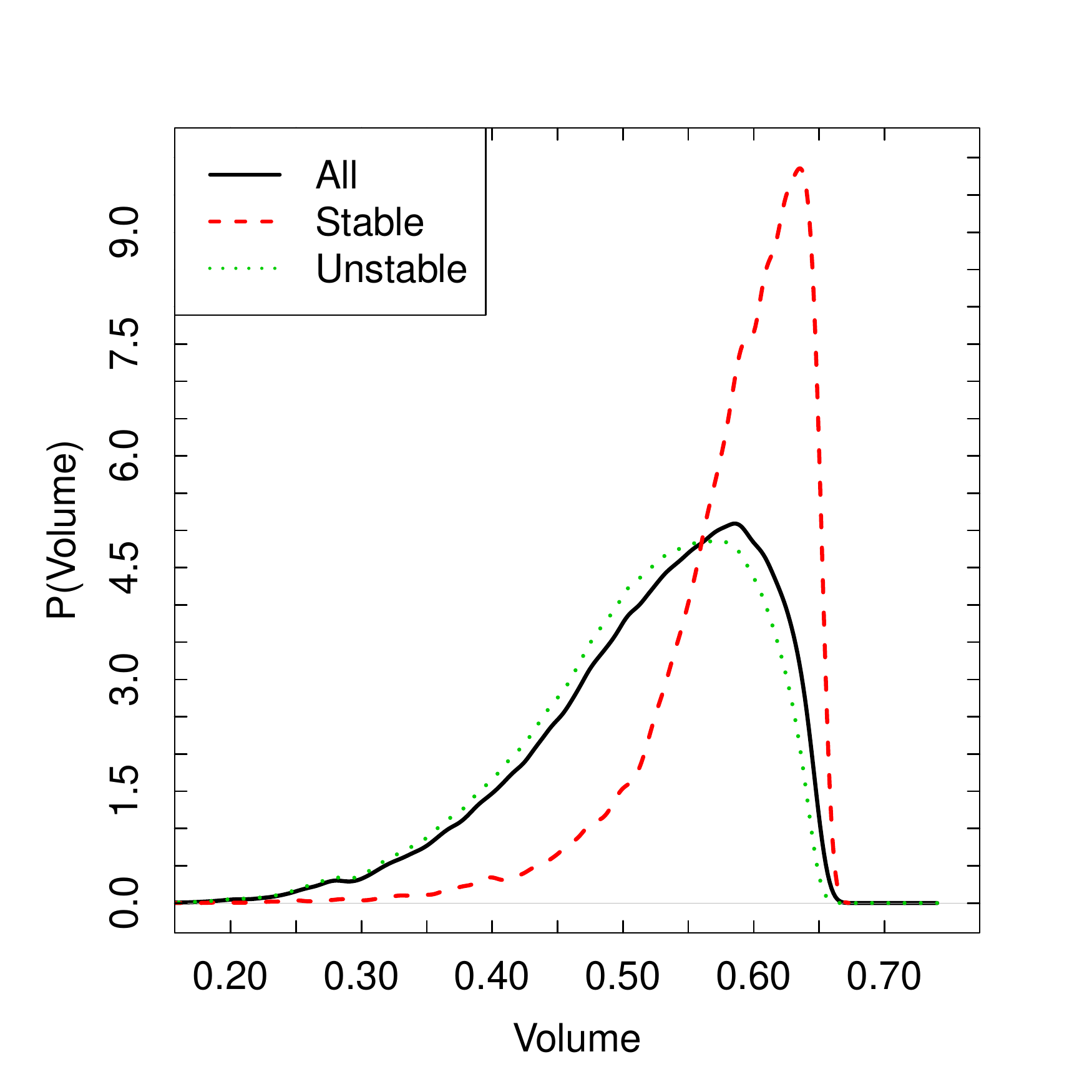}
  \caption{The distribution of volumes (black) as sampled over a grid of $2^{20}$ points
    spanning the phase space. The volume spectrum of the stable regions is shown as the dashed red curve,
    the unstable regions are shown by the dashed green curve. Note that the area of all curves shown are normalized to one.}
  \label{fig:volume-spectrum}
\end{figure}

The distribution of stable regions is generally correlated with larger values of $p_1,
p_2$ suggesting larger volumes. In \figref{fig:volume-spectrum} we show the distribution
of volumes for stable and unstable regions, stable regions are much more likely to be
associated with higher volumes. The mode of the stable curve is at $V = 0.645$, the mode
of the unstable curve is at $V=0.587$. The volume of an equilateral triangular prism with
unit face areas is $\frac{3^{1/4}}{2} = 0.658$, stable regions are associated with more
regular polyhedral configurations. As the figure shows this relationship is not cut and
dried as the high mode of the unstable curve shows there is a sizable fraction of unstable
configurations at relatively large volumes.

\begin{figure}[t]
  \centering
  \includegraphics[width=0.35\textwidth]{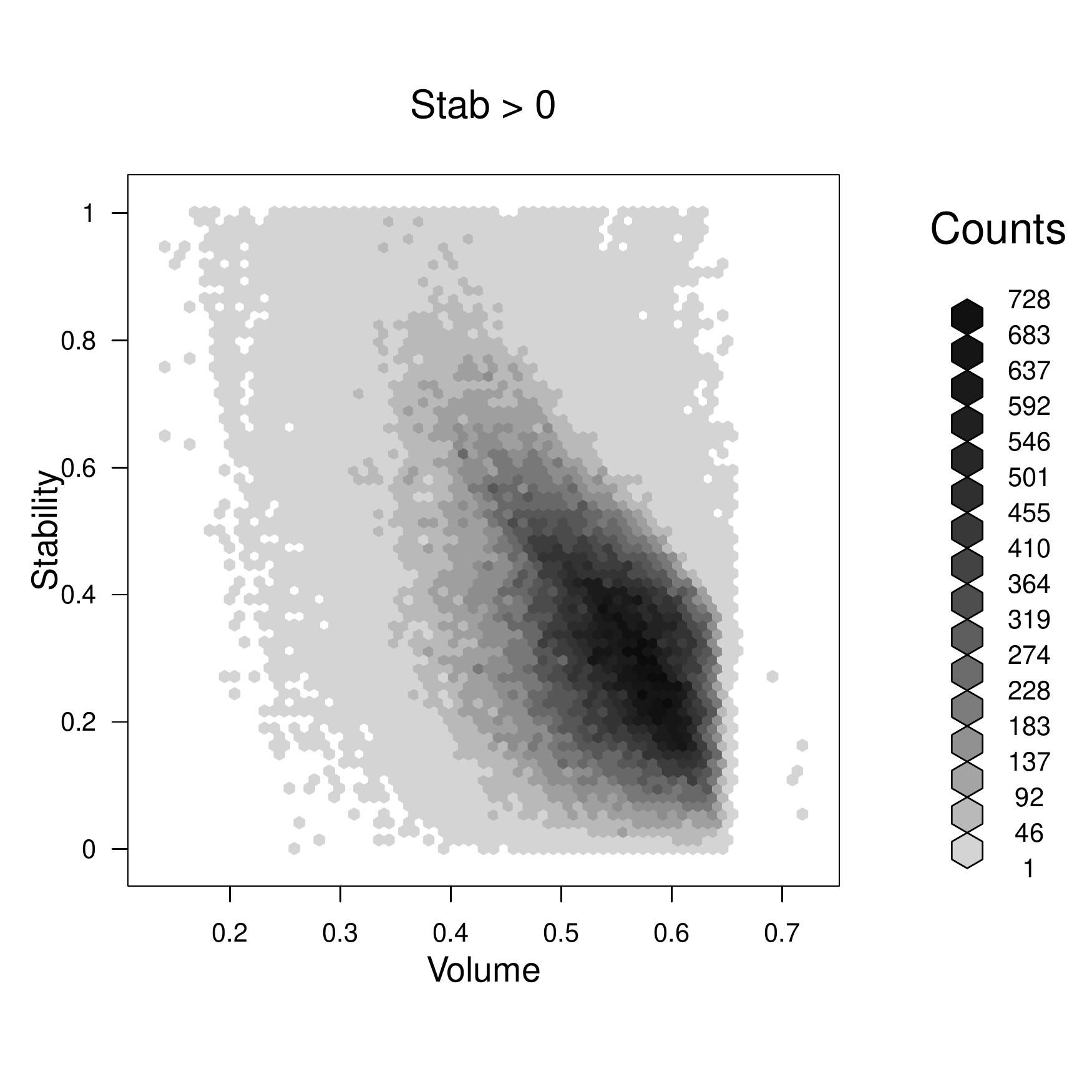}
  \caption{The density of the positive real components of LLE's plotted against the volume of the system, the
    LLE's are restricted to the nearly stable region ($0 < \max\{\Re(\lambda_{z}^{+})\} < 1$ ).}
  \label{fig:density-stable}
\end{figure}

The volume distribution for almost stable regions $0 < \max\{\Re(\lambda_{z}^{+})\} < 1$
is plotted against their stability in \figref{fig:density-stable}. There is a clear linear
correlation between volume and stability in the highest density region of the figure. This
correlation explains the relatively high peak of the inclusive and unstable curves in
\figref{fig:volume-spectrum}, these are associated with weakly unstable regions in the
phase space.  The majority of stable, or nearly stable, configurations correspond to those
with a large volume.

Knowing the local stability of a point is most useful if we have some measure of how
likely it is that the a trajectory of the system will actually pass through this
point. Ideally this measure would be computed by dividing the phase space into small
volumes, integrating many trajectories, and then calculating the probability for a
trajectory to reach each of these points. The Kolmogorov-Sinai entropy arises directly from
this construction (cite). As this is currently numerically untenable we seek an alternative
measure for the probability of the system being in a given configuration.

\subsection{Canonical Stability Estimates}
We consider the system as a member of a canonical ensemble, i.e we embed the system into a
heat bath at some fixed temperature $T=\beta^{-1}$.  This non-isolated limit is a reasonable choice
if our polyhedron is part of some extended interacting system of polyhedra.
In the canonical ensemble we can assign the following weight to any particular
configuration $\sigma$ of the system
\begin{equation}
  \label{eqn:canonical-measure}
P(\sigma) = \begin{cases} 
  \exp(- \beta H(\sigma)) \quad &\mbox{if } H(\sigma) > 0 \\
  0 \quad &\mbox{if $H(\sigma)$ undefined} ,
  \end{cases}
\end{equation}
where the second case explicitly assign zero weight to points where the Hamiltonian cannot be
evaluated or where the numerical routines fail. The Kolmogorov-Sinai (KS) entropy for a system
can be computed from the sum of positive GLE's, the largest positive LLE can serve as
local estimate of this. The eigenvalues of $\mathcal{H}$ are opposite signed pairs, in
unstable regions the secondary positive eigenvalue is usually small so this is a
reasonable approximation. Using the canonical measure \eqref{eqn:canonical-measure} 
we propose the following canonical instability measure as an estimate of the KS entropy
\begin{equation}
  \label{eqn:ks-ent}
  \hat{h}(T) = \frac{\int_{\mathcal{U}} \exp(- \beta H(\sigma)) \max\left\{\Re(\lambda_{z}^{+})\right\} d\sigma}{\int_{\mathcal{U}} \exp(- \beta H(\sigma)) d\sigma}
\end{equation}
where the region $\mathcal{U}$ is the set of unstable points in phase space,
i.~e.\ points where $\max\{\Re(\lambda_z^{+})\} > 0$. 
In \tabref{tab:ks-results} we compare estimates for the KS entropy along with
the fractions of stable and unstable volumes of the phase space for the unit area
triangular prism, the Henon-Heiles (HH) potential $V(x,y) = \frac{1}{2}(x^2+y^2+2x^2y - 2/3y^3)$ 
and the two-dimensional classical Yang-Mills (CYM) potential $V(x,y) = \frac{1}{2}x^2y^2$. 
The latter are both well known non-integrable systems which display aspects of 
Hamiltonian chaos, see \cite{Muller:1994, Licht:1992, Gutzwiller:1990} and references therein 
for more details. As the Table shows, the triangular prism has the largest relative volume of local
stability, being almost twice that of the CYM and HH systems. The HH system is known to be
non-integrable for larger energies, the CYM system is non-integrable for all energies. 
That our estimates of the KS entropy for the triangular prism is of a similar magnitude is 
strongly suggestive that it is also non-integrable and exhibits some degree of mixing.

\begin{table*}
  \begin{ruledtabular}
    \begin{tabular}{l c c c c c}
    Name & Momentum Range & Position Range & Stable Fraction & Unstable Faction & $\hat{h}(1) $ \\
    \hline 
    Henon-Heiles &  $[-2, 2] \times [-2,2]$ & $[-2,2] \times [-2,2]$ & 0.06 & 0.94 & 1.36 \\
    Classical Yang Mills &  $[-2, 2] \times [-2,2]$ & $[-2,2] \times [-2,2]$ & 0.08 & 0.921 & 0.548 \\
    Triangular Prism &  $\approx [0, 2] \times [0,2]$ & $[0,2\pi] \times [0,2\pi]$ & 0.13 & 0.86 & 0.38 \\
  \end{tabular}
  \end{ruledtabular}
  \caption{\label{tab:ks-results}The fraction of stable and unstable regions along with
    the canonical stability measure \eqref{eqn:ks-ent} for the triangular prism system (with
    all faces set to 1), the Henon-Heiles potential and the two-dimensional classical
    Yang-Mills potential. The temperature is chosen as $T=1$.}
\end{table*}

\subsection{Microcanonical Stability Estimates}  
To examine the energy (volume) dependence of the instability measure we consider the 
microcanonical ensemble. Estimates using the microcanonical ensemble are valid for 
isolated systems. If the system is ergodic then microcanonical averages are equivalent to
averages over trajectories in the long time limit. The microcanonical density $\Omega(E)$ is 
\begin{equation}
  \label{eqn:micro-canonical-density}
  \Omega(E) = \int_{V} \delta(E - H(p,q))\;dp dq,
\end{equation}
where the integral is computed over the entire phase space. The total volume distribution,
the black curve, shown in \figref{fig:volume-spectrum} is proportional to the
microcanonical density for the equal-areas triangular prism.

\begin{figure}[ht]
  \centering
  \includegraphics[width=0.4\textwidth]{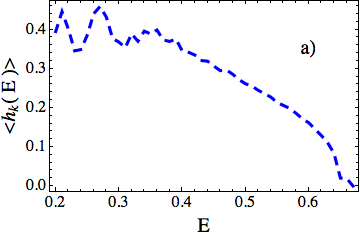}
  \hspace{0.2cm}
  \includegraphics[width=0.4\textwidth]{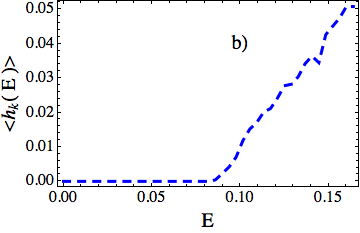}
  \caption{Microcanonical instability measure $h_k(E)$, computed using LLE's, for
    the equal area triangular prism (a) and a particle moving in the HH potential (b).}
  \label{fig:ks-ent-micro}
\end{figure}

The associated microcanonical proxy for the KS entropy is 
\begin{equation}
  \label{eqn:micro-canonical-ent}
  \langle h_k(E) \rangle = \frac{1}{\Omega(E)} \int_{V} \max\left\{\Re(\lambda_{z}^{+})\right\} \delta(E - H(p,q))\; dp dq.
\end{equation}
In \figref{fig:ks-ent-micro} we plot the microcanonical instability measure computed from LLE's 
for the unit area triangular prism and for a free particle in the Henon Heiles potential as 
function of the energy $E$. The instability measure of the triangular prism is high at small 
energies (pentahedron volumes) and decreases rapidly to zero at large energy. 
Small volume configurations exhibit a high degree of instability, because the system
flops between the many available equivalent configurations. At larger volumes the set of
possible volume preserving deformations is reduced and the instability decreases. As a
comparison we have also computed the microcanonical instability measure for the HH system 
using \eqref{eqn:micro-canonical-ent}. The HH system is stable for $E < 0.1$, and the instability
rises rapidly for for larger energies. At low energies the system is known to be integrable 
\cite{Licht:1992}. 

\subsection{Intermediate Lyapunov Exponents}

Given the promising results of the local Lyapunov analysis we turn to examining
intermediate Lyapunov exponents (ILE's). The LLE's suggest the existence of large areas of
local instability however there may be some complicated dynamical conspiracy that allows
for the existence of stable periodic trajectories. The ILE's provide a bridge between the
LLE's, which are the zero time limit of the ILE's, and the global Lyapunov exponents
(GLE's) which are the infinite time limit of the intermediate Lyapunov exponents.

We compute the maximum positive ILE by the method of Benettin et al
\cite{Benettin:1976dy}. A reference trajectory is numerically integrated for some period
of time $z(t)$. A vector $d$ is generated by some small deviation $d_0$ from the
starting point of the reference trajectory which is then also integrated. The largest
positive intermediate Lyapunov exponent after some time $t_{n}$ is
\begin{equation}
  \label{eqn:intermediate-lyapuov}
  \lambda_n = \frac{1}{t_{n} - t_0} \log \frac{d(t_{n})}{d_0}.
\end{equation}
In practice the distances between the trajectories $d(t)$ grow rapidly, to improve
numerical stability the distances are periodically rescaled once they exceed some
fixed threshold $D$. We define a rescaling 
\[
\alpha_1 = \frac{d(t_1)}{d(t_0)} \quad \{t_1\; |\; d(t_1 ) > D\},
\]
repeating this process as required the ILE \eqref{eqn:intermediate-lyapuov} can be written as
\begin{equation}
  \label{eqn:intermediate-lyapuov-2}
  \lambda_{n} =\frac{1}{t_{n}-t_0} \sum_{i=1}^{n} \log \alpha_i,
\end{equation}
where $\alpha_i = \frac{d(t_i)}{d(t_0)}$. 

The reference and deviation trajectories were integrated using an implicit symplectic
Runge-Kutta (RK) integrator \cite{Hairer:2000}. A computationally more expensive implicit
method was chosen as the Hamiltonian cannot be readily separated into momentum and
potential components making explicit RK splitting methods untenable. The implicit
equations are solved iteratively. The iterations are initialized using an equistage
method. Symplectic integrators explicitly preserve the two-form $\Omega = d \vec{q} \wedge
d\vec{p}$ upon the phase space. A symplectic integrator generates maps which are
themselves symplectic, orbits of the system under this map can be thought of as following
a shadow solution to those of the exact Hamiltonian. Orbits under non-symplectic maps may
follow the exact Hamiltonian more closely for small times but in general will diverge from
the true dynamics over long time scales \cite{Hairer:2000, Leimkuhler:2004}. A fourth
order Gauss collocation method was found to provide a satisfactory balance of accuracy and
computational efficiency \cite{Hairer:2000,Butcher:1964a}. The stability properties of these
methods ensure that the dynamics does not depend upon the choice of time step for the
integrator, the numerical convergence of the iterative process is slower at larger time
steps.

A large set of isochoric points were sampled from the Hamiltonian and used to generate
trajectories and ILE's, this process was carried out for a set of volumes $\{0.35, 0.45,
0.55, 0.65\}$ so that the energy dependence of the ILE's could be computed.

\begin{figure}[ht]
  \centering
  \includegraphics[width=0.5\textwidth]{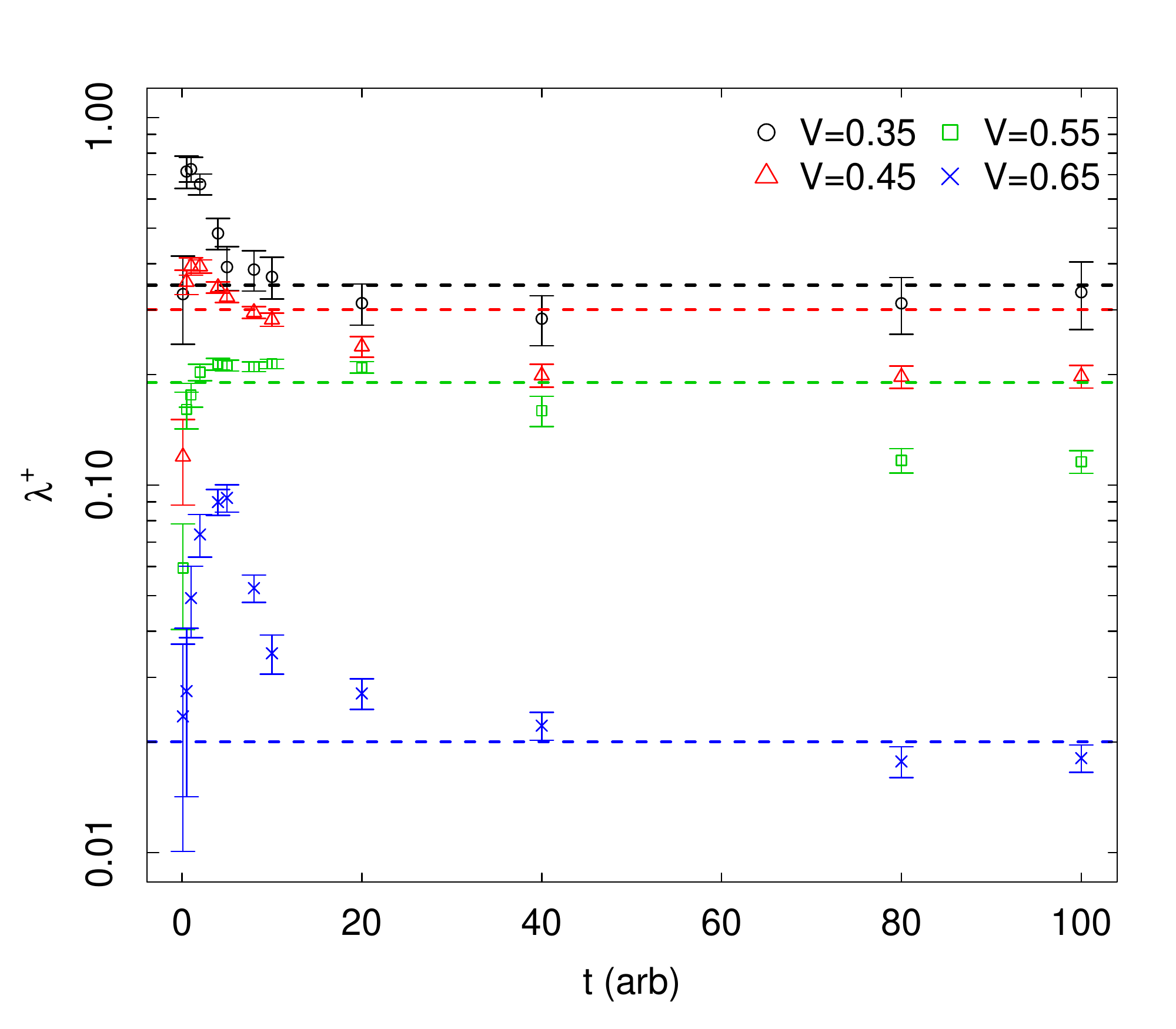}
  \caption{The ensemble averaged intermediate Lyapunov exponents (ILE) computed for
    ensembles of trajectories with volume $\{0.35, 0.45, 0.55, 0.65\}$, the bars show
    standard errors. The dashed horizontal lines show the appropriate microcanonical
    estimates of the LLE's. }
  \label{fig:ile-time}
\end{figure}

\begin{figure*}[hb]
  \centering
  \includegraphics[width=0.6\textwidth]{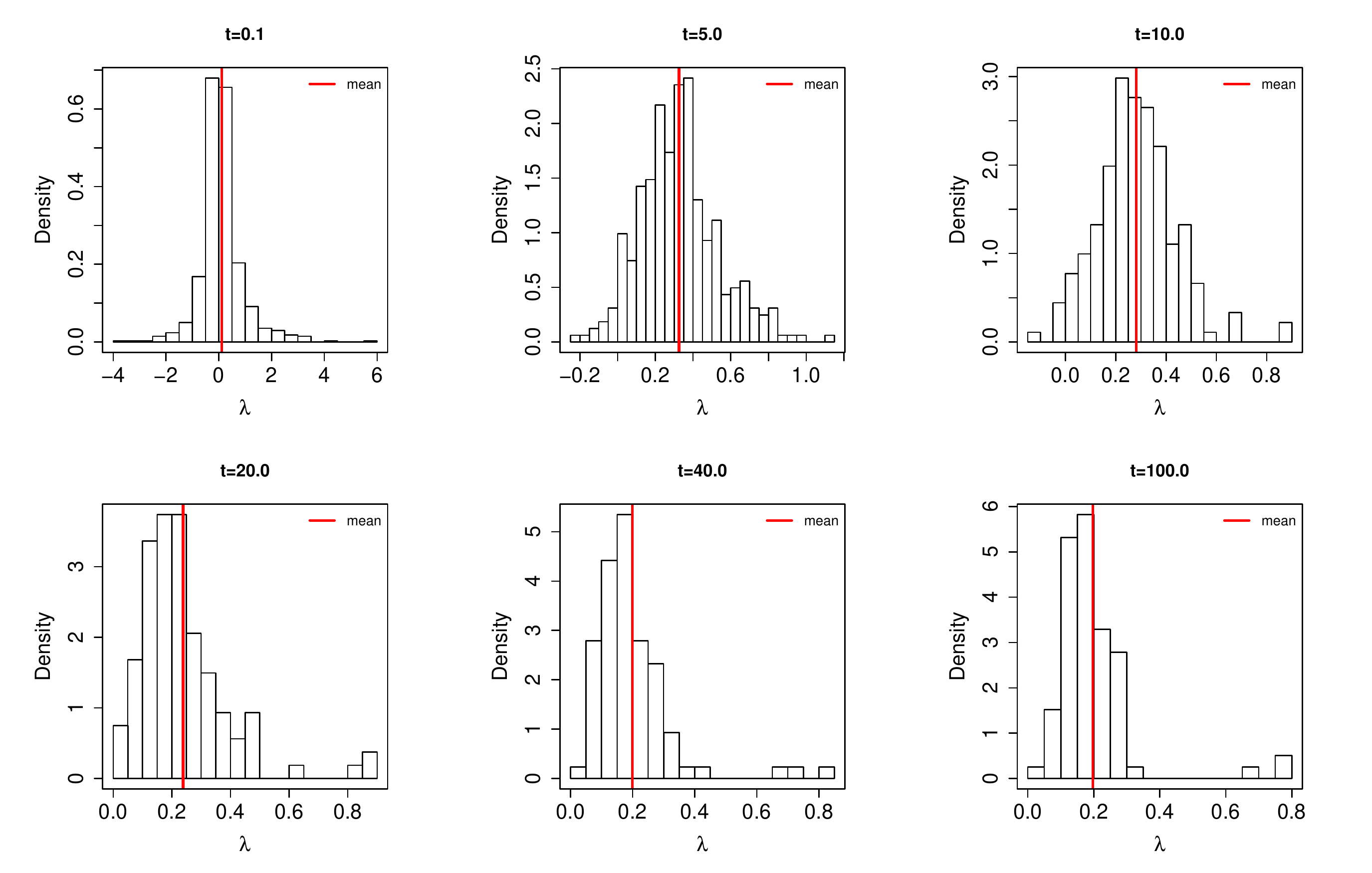}
  \caption{The distribution of intermediate Lyapunov exponents (ILE) for trajectories
    with $V=0.45$ plotted at increasing times $t=\{0.1, 5, 10, 20, 40, 100\}$. The vertical
    red line shows the mean of the distribution.}
  \label{fig:ile-distribution}
\end{figure*}

The time evolution of the ensemble average of the ILE's for each of the volumes is shown
in \figref{fig:ile-time}, note that the y--axis is log scaled. The dashed lines show the
microcanonical LLE values computed from \eqref{eqn:micro-canonical-ent}. The ILE's rapidly
peak and then slowly saturate to stable long time values. Note that the rise time is
different for the different volumes, the more unstable small volume ($V=0.35$) ILE's grow
faster than the more stable high volume case. In \figref{fig:ile-distribution} the time evolution of the distribution of
ILE's generated for the $V=0.45$ case is shown. The distribution is centered around zero
for small times, the mean converges to $0.2$ by $t=40$. The small time limit does not
reproduce the microcanonical estimate of the LLE's. For very short times the LLE's have
more information about the local instability of the system than the ILE's as they are
generated from the full Jacobian of the Hamiltonian gradient. The LLE's tend to fluctuate
around zero for small times as they are generated from random initial displacements which
may be initially along a stable direction. As we observe in \figref{fig:ile-time} over
time the most unstable direction will dominate and the distributions appear to converge. 

The long time values should be reasonable approximations to the GLE's. The ILE's are
larger at low volume and become very small at $V=0.65$, this matches well with the
microcanonical estimates. The unit pentahedron is strongly chaotic at small volumes
and becomes more regular as the volume approaches it maximum limit.

\section{Conclusions}
\label{sec:conclusion}
In our investigation of the phase space of the unit area triangular prism we have found a
great deal of structure in the Hamiltonian and in the distribution of configurations. The
phase space contains moderate regions of local stability and large regions of local
dynamical instability. The distribution of local Lyapunov exponents appears to be correlated 
with the boundaries in the configuration space. We have calculated the average dynamical
instability measures in the canonical and microcanonical ensembles and obtained values 
that are comparable to those found in well-known chaotic systems.  The triangular prism 
differs from these and most other dynamical systems in that the stability increases with energy. 
Higher energy (high volume) triangular prisms are dynamically more stable than low energy 
(low volume) prisms. 

The large degree of dynamical instability found
in our investigation of the isochoric pentahedron with unit area faces provides an encouraging 
starting point for a bottom-up investigation of the origin of thermal behavior of gravitational field 
configurations in loop quantum gravity. That the dynamical instability occurs in the simplest
polyhedron where it can suggests that it will be a generic property of more complex polyhedra. 
Any coupling to other polyhedral configurations can be expected to enhance the degree of 
instability. We emphasize again the most interesting result of our investigation, that the
rate of instability does not go to zero at small energy. In other words, the low energy dynamics
of the isochoric pentahedron is not characterized by ballistic trajectories in phase space, but by 
rapid and energy independent diffusion among different degrees of freedom. At low energies,
the pentahedron appears to be a {\em fast scrambler} of information.

It is tempting to speculate whether this unusual 
property has implications for the quantum theory of gravity. Dynamical chaos in classical field 
theories is related to the instability of the perturbative effective action of the quantum field 
theory \cite{Matinyan:1997rd,Matinyan:1996hd}. Because the full effective action is known
to be convex, and thus dynamically stable, for any quantum field theory \cite{Fujimoto:1982tc},
this instability is usually cured by some dynamical mechanism, such as mass generation 
through spontaneous symmetry breaking or confinement, or the breaking of translational 
invariance. None of these mechanisms are realized in the case of gravity, which is unique
in being characterized by the simultaneous absence of an infrared mass scale and the
existence of a ultraviolet scale, the Planck mass, which controls the quantum corrections
to its classical limit.

Returning to the dynamics of the pentahedron, we intend to report on the systematic
numerical integration of a phase-space spanning set of trajectories in a future work. The
preliminary efforts required to compute the ILE's have identified small regions of
quasi-periodicity. The isochoric pentahedron thus does not exhibit full ergodicity. The
isochoric pentahedron being a dynamical system with rich and fascinating dynamics, larger
polyhedra with $N>5$ will likely be even more so. Investigation of the $N=6$ system would
be particularly interesting, as it would expose an additional aspect in the structure of
the phase space as the hexahedron passes between the two dominant classes of six-faced
polyhedra.

The non-integrability of the classical triangular prism indicated by our analysis rules out 
a straightforward application of Bohr-Sommerfeld quantization, which was successfully used
to explore the quantum mechanics of the tetrahedron \cite{Bianchi:2011ub}. Methods from the 
study of quantum chaos such as the Gutzwiller trace formula \cite{Gutzwiller:1990} may provide 
some insight into the quantum behavior of the pentahedron and higher polyhedra. Variational 
methods or diagonalization in some suitable basis may also prove fruitful.

\begin{acknowledgments} 
  We acknowledge support by DOE grant DE-FG02-05ER41367. This research was done 
  using computing resources provided by OSG EngageVo funded by NSF award 075335. 
  CCS would like to thank H.~Haggard, J.~Powell and particularly I.~C.~Kozyrkov for many
  helpful discussions. BM thanks E.~Bianchi for useful discussions.
\end{acknowledgments}

\bibliographystyle{apsrev} 
\bibliography{./qgrav-refs}{}

\end{document}